\documentclass[% 
 amsmath,amssymb,
 aps,
]{revtex4-1}
\usepackage{graphicx}
\usepackage{dcolumn}
\usepackage{bm}
\usepackage{epstopdf}

\begin{document}

\preprint{APS/123-QED}

\title{ Algebraic cluster models calculations for shape phase transitions of boson-fermion systems}

\author{ M.Ghapanvari}
\email{m.ghapanvari@tabrizu.ac.ir}
\affiliation{Department of Nuclear Physics, University of Tabriz, Tabriz 51664, Iran.}

\author{ N.Amiri}
\email{narjesamiri@tabrizu.ac.ir}
\affiliation{Department of Nuclear Physics, University of Tabriz, Tabriz 51664, Iran.}

\author{ M. A. Jafarizadeh}
\email{jafarizadeh@tabrizu.ac.ir}
\affiliation{Department of Theoretical Physics and Astrophysics,
University of Tabriz, Tabriz 51664, Iran.}
\affiliation{Research Institute for Fundamental Sciences, Tabriz 51664, Iran}

\date{\today}

\begin{abstract}
The Algebraic Cluster Model(ACM) is an interacting boson model that gives the relative motion of the cluster configurations in which all vibrational and rotational degrees of freedom are present from the outset. We schemed a solvable extended transitional Hamiltonian based on affine $ {{SU(1,1)}} $ Lie algebra within the framework for two-, three- and four- body algebraic cluster models that explains both regions $ O(4)\leftrightarrow U(3) $, $ O(7)\leftrightarrow U(6) $ and $ O(10)\leftrightarrow U(9) $, respectively . 
We offer that this method can be used to study of $k\alpha + x$ nucleon structures with k = 2, 3,4 and x = 1, 2, . . . , in specific x = 1,2 such as structures $^{9}Be$,$^{9}B$,$^{10}B$ ; $^{13}C$, $^{13}N$, $^{14}N$; $^{17}O$, $^{17}F$.  
Numerical extraction to the energy levels, the expectation value of boson number operator and behavior of the overlap of the ground-state wave function within the control parameters of this evaluated Hamiltonian are presented. 
The effect of the coupling of the odd particle to an even-even boson core is discussed along the shape transition and, in particular, at the critical point.

\end{abstract}

\maketitle
\section{  1.Introduction}

Algebraic models are advantageous in the many-body and in few-body systems. In  algebraic models energy eigenvalues and eigenvectors are obtained by diagonalizing a finite-dimensional matrix, rather than by solving a set of coupled differential equations in coordinate space. As an example, we  assign the interacting boson model (IBM), which has been very prosperous in the appositive of the collective states in nuclei \cite{1}. Its dynamical symmetries correspond to the quadrupole vibrator, the axially symmetric rotor and the $ \gamma $-unstable rotor in a geometrical description. In addition to these special solutions, the IBM can describe the intermediate cases between any of them equally well. The first application of the algebraic approach to the few-body systems was the vibron model \cite{2}, which was recommended to describe the vibrational and rotational excitations in the diatomic molecules.
Algebraic methods give an accurate view  to spectroscopic studies and focus on  their's symmetries and selection rules to categorize  the basis states, and to evaluate   matrix elements of physical observables \cite{3}. The binding energy per nucleon for the light nuclei shows large oscillations with the nucleon number with maxima for nuclei with A =4n and Z=N, especially for the nuclei $\mathrm{^{4}He}$, $\mathrm{^{8}Be}$, $\mathrm{^{12}C}$ and $\mathrm{^{16}O}$ for n =1, 2, 3 and 4 respectively, which provides a strong indication of the importance of $  \alpha  $ clustering in these nuclei \cite{4}. The common method is to introduce a $  U(\nu +1) $ spectrum generating algebra for a bound-state problem with $ \nu  $ degrees of freedom in which all states are assigned to the symmetric representation $ [N] $ of $ U(\nu + 1) $ \cite{5,6}. For the $  \nu = 5 $ quadrupole degrees of freedom in collective nuclei this led to the introduction of the $  U(6)  $ interacting boson model \cite{1}. Similarly, the $ U(4) $ vibron model was proposed to describe the dynamics of the $ \nu = 3 $ dipole degrees of freedom of the relative motion of the two objects, e.g. two atoms in a diatomic molecule \cite{2}, the two clusters in a nuclear cluster model \cite{7,8,9}, or a quark and antiquark in a meson \cite{10,11}. An application to the three-body system involves the six degrees of freedom of the two relative vectors which in the algebraic approach leads to a $ U(7) $ spectrum generating algebra  \cite{12,13}as an extension of the vibron model. The $ U(7) $ model was developed originally to describe the relative motion of the three constituent quarks in baryons \cite{12,13}, but it has also found applications in molecular physics \cite{14,15} and nuclear physics ($\mathrm{^{12}C}$ as a cluster of three a particles) \cite{4,5}. The algebraic cluster for the four-body systems in terms of a $U(10) $ spectrum-generation algebra was introduced in  \cite{15}. An application to the cluster states in $\mathrm{^{16}O}$ suggested that these can be interpreted in terms of rotations and vibrations of tetrahedral configuration of α particles. The triangular configuration in $\mathrm{^{12}C}$ and tetrahedral configuration in $\mathrm{^{16}O}$ implied by the observed rotational sequence, were confirmed by a study of BE(L) electric transitions along the ground state bands  \cite{4,5,16}. In \cite{17} $\mathrm{^{8}Be}$, $\mathrm{^{12}C}$ and $\mathrm{^{16}O}$ nuclei were considered by using an infinite-dimensional algebraic method based on the affine $ {{SU(1,1)}} $ Lie algebra for the transitional descriptions of the vibron model and $ \alpha $-cluster model. The cluster structures with addition of nucleons discussed especially in  the Be isotopes with a variety of methods \cite{17,18,19,20,21,22,23,24,25,26,27,28,29,30}. In Ref \cite{31,32}, single-particle levels in cluster potentials in $k\alpha + x$ nucleon structures within the framework of a cluster shell model (CSM) calculated. In nuclear physics s-orbit and p-orbit adjaceny achived by studying  in light nuclei as we see  in carbon isotopes \cite{18}. Hafstad and Teller studied (4n + 1) nuclei, e.g. $^{9}Be$, $^{13}C$ and $^{17}O$. Their ideas were based upon the structure of the $^{5}He$ nucleus in which the last neutron was in a p-orbit \cite{18,21,22}.

The study of the quantum phase transitions enjoys a substantial interest in the algebraic models of the nuclear structure. There is mutual relations between shapes (phases) and dynamic symmetry limits.  The  analytical solutions provide a process in which the system undergoes a change from one dynamical symmetry to another one. The first examples \cite{33,34} were related to the Interaction Boson Model Approximation (IBA) \cite{1} and the Vibron model \cite{1,3}.
 
The aim of this contribution is to discuss the quantum phase transitions in the algebraic cluster models for the two-, three- and four- body cluster, to transition description in $ U(3)\leftrightarrow O(4) $, $ U(6)\leftrightarrow O(7) $ and $ U(9)\leftrightarrow O(10) $. This model can be solved by using an infinite dimensional algebraic technique in the IBM framework.  This method was applied to the $k\alpha + x$ nucleon structures consisting of $ k $ $ \alpha $-particles and $ x $ nucleons, such as structures $^{9}Be$,$^{9}B$,$^{10}B$ ; $^{13}C$, $^{13}N$, $^{14}N$; $^{17}O$ and $^{17}F$, corresponding to the exchange of neutrons and $\alpha$-particles. In order to describe the phase transition, we calculate some observables such as energy level, level crossing, expectation values of boson number operator and  overlap of the ground-state wave function. The results of calculations for these nuclei are presented and are compared with the corresponding experimental data.
In this work ,the role of a fermion with angular momentum j at the critical point on quantum phase transitions in bosonic systems is investigated.

The specific aims of the present study and the structure of this paper are as follows:
In section 2, we introduce the algebraic cluster model, followed by a discussion of the permutation symmetry. Section 3 briefly summarizes the theoretical aspects of the model. Numerical results are presented in section 4 and section 5 is devoted to summarize and to justify some conclusion.

  \section{The algebraic cluster model (ACM)}
   
   In this section, we introduce the algebraic cluster model. It is based on the spectrum generating algebra of $U(\nu+1)$, where $\nu = 3(n-1)$ represents the number of the relative spatial degrees of freedom. As special cases the ACM contains the $U(4)$ vibron model \cite{2} for the two-body problems $(n=2)$, the $U(7)$ model \cite{12,13,15,4,5} for three-body clusters $(n=3)$ and the $U(10)$ model \cite{16,35,36} for four-body clusters $(n=4)$.
   
   The relevant degrees of freedom of a system of n-body clusters are given by the $ n-1 $ relative Jacobi coordinates

   \begin{equation}
  \vec{\rho}_k=\frac{1}{\sqrt{k(k+1)}}(\sum_{i=1}^{k}\vec{r}_i -k\vec{r}_{k+1} )  \quad  \quad k=1,2,...,n-1  
   \end{equation}
   and their conjugate momenta. Here $ \vec{r}_i $  denotes the position vector of the $ i-th $ cluster (i=1,2,...,n). 
   
   Instead of a formulation in terms of coordinates and momenta, the method of bosonic quantization is used which consists of introducing a dipole boson with $ L_{P}  =1^{-} $ for each independent relative coordinate and an auxiliary scalar boson with $ L_{P}  =0^{+} $
      \begin{equation}
    s^{\dagger}  ,b_{k,m}^{\dagger}            
      \end{equation}
 with $ k = 1,...,n-1 $ and $ m = −1,0,1 $. The scalar boson does not represent an independent degree of freedom, but is added under the restriction that the Hamiltonian commutes with the number operator 
     \begin{equation}
   N=  s^{\dagger}s+ \sum_{k}\sum_{m}b_{k,m}^{\dagger}b_{k,m}            
       \end{equation} 
   i.e. the total number of bosons $  N=n_s+\sum_{k}n_k $ is conserved. The set of $ 〖(3n-2)〗^2 $ bilinear products of creation and annihilation operators spans the Lie algebra of $ U(3n-2) $.
   
   In this contribution, we study the ACM for identical clusters which is relevant to $ \alpha $-cluster nuclei such as $\mathrm{^{8}Be}$, $\mathrm{^{12}C}$ and $\mathrm{^{16}O}$. For these systems, the Hamiltonian has to be invariant under the permutation group $ S_n $. The permutation symmetry of n identical objects is determined by the transposition $ P(12) $ and the cyclic permutation $ P(12...n) $ \cite{37}. All other permutations can be expressed in terms of these two elementary ones. The transformation properties under $ S_n $ of all operators in the model originate from those of the building blocks. The scalar boson, $  s^{\dagger} $, transforms as the symmetric representation $ [n] $, whereas the dipole Jacobi bosons, $ b_k^{\dagger} $ with $ k = 1,...,n-1 $ transform as the $ n-1 $ components of the mixed symmetry representation $ [n-1,1] $.
  
  Hamiltonian that describes the relative motion of a system of n identical clusters, and is a scalar under the permutation group $ S_n $ and is rotationally invariant. It conserves the parity as well as the total number of bosons, as given by
  
      \begin{equation}
      H=\epsilon_0 s^{\dagger} \tilde{s}-\epsilon_1\sum_{i}b^{\dagger}\tilde{b}+u_0 s^{\dagger} s^{\dagger} \tilde{s} \tilde{s}-u_1 \sum_{k}s^{\dagger}b_{k}^{\dagger}s^{\dagger}\tilde{b}_{k}+\nu_0 (\sum_{k}b_{k}^{\dagger}\tilde{b}_{k}\tilde{s}\tilde{s}+hc)+\sum_{L}\sum_{ijkl}\nu^{(L)}_{ijkl}[b_{i}^{\dagger}\times b_{j}^{\dagger}]^{(L)}[\tilde{b}_{k}\times \tilde{b}_{l}]^{(L)} 
      \end{equation} 
      
      with $ \tilde{b}_{k,m}={(-1)}^{(1-m)} b_{k,-m}  $ and $ \tilde{s}=s $ by construction, the $ \epsilon_0 $, $ \epsilon_1 $, $ u_0 $, $ u_1 $ and $ \nu_0 $ terms in Equation (4) are invariant under $ S_n $.
      
      In this contribution, we consider two dynamical symmetries of the ACM Hamiltonian for the n-body problem which are related to the group lattice
      
    \begin{eqnarray}        
         U(3n-2)\supset\{^{U(3n-3)}_{O(3n-2)}\supset O(3n-3) \nonumber
    \end{eqnarray}
          
          which are called the $ U(3n-3)$ and $SO(3n-2)$ limits of the ACM, respectively. A geometric analysis shows that the $U(3n-3)$ limit corresponds for large N to the (an)harmonic oscillator in $3(n-1)$ dimensions and the $SO(3n-2)$ limit to the deformed oscillator in $3(n-1)$ dimensions  \cite{35,36} .

\section{Theoretical framework}

   Algebraic models provide elegant and simple paradigms for the behavior of a wide variety of physical systems. The basic idea of algebraic models is that Hamiltonians and other physical operators of these systems can be realized by using a set of boson operators, since the collective excitations of these systems can be regarded as a set of interacting bosons. The spectrum of the systems can be generated by an appropriate unitary Lie algebra, called spectrum generating algebra. Dynamical symmetries often play an important role in the approach. There is one to one correspondence between the shapes (phases) and dynamic symmetry limits in which analytical solutions to the model exist. The shape (phase) transition of these models is referred to as a process in which the system undergoes a change from one dynamical symmetry to another one. The method for diagonalization of the Hamiltonian in the transitional region is not as easy as in either of the limits, especially when the dimension of the configuration space is relatively large. To avoid these problems, an algebraic Bethe ansatz method within the framework of an infinite dimensional $ SU(1,1) $ Lie algebra has been proposed by Pan et al... \cite{38,39}.

   \subsection{The $ SU(1,1) $ expression of Bethe ansatz equations for two-cluster systems}
                                           
 \subsubsection{The odd-A nuclei : $ ^{9}_{4}Be$ and $ ^{9}_{4}B$} 
For more than two decades, is known that the  $ ^{9}_{4}Be$ nucleus  is an example of a molecular covalent bond in nuclear physics, Where two particles with valence neutron  are limited.  
The  $ ^{9}_{4}Be$ nucleus , which has unlimited system properties $2\alpha+n$ is the
“cornerstone” of cluster physics\cite{18,21,22,32}.
Due to its low neutron separation threshold, separation of $ ^{9}_{4}Be$ can be an origination of astable $ ^{8}_{4}Be$ nuclei.
The  $ ^{8}_{4}Be$ isotope is known as
the only nucleus whose ground state is distinguished as the $\alpha$-particle Bose condensate.
A study of the division of the $ ^{9}_{4}Be$ nucleus in $\alpha$-particle pair appears to be a clear starting point than the more complex $N\alpha$-systems. This method can also be used to describe the odd-A nuclei. For example,  $ ^{9}_{4}Be$  may be assumed to be composed of two $\alpha$-particles and a valence neutron, forming, at larger
$\alpha+\alpha$ separations $ ^{5}He$ nuclei, where the neutron abides in a $p_{3/2}$-orbit. We assume for $ ^{9}Be$ a structure similar to $ ^{9}Be$, with the odd neutron exchanged by an odd-proton\cite{18,21,22,32}.
  
The boson algebraic structure will be always taken to be $\mathrm{U^{B}(4)}$ , while the fermion algebraic structure will depend on the values of the angular momenta, j, taken into consideration \cite{10}.
Two possible dynamical symmetry limits, $ U^{B}(3) $ and $ O^{B}(4) $, are related to the following two algebraic chains,
 \begin{equation}
  U^{B}(4)\otimes U^{F}(2j+1)\supset\left\{ \begin{array}{l}
{U^{(B)}}(3)\\
{O^{(B)}}(4)
\end{array} \right\}\
\otimes SU^{F}(2j+1)\supset O^{B}(3)\otimes Sp^{F}(2j+1) \supset O^{B}(3)\otimes SU^{F}(2) \supset  Spin^{BF}(3) 
 \end{equation}  
The negative parity states in the odd-mass nuclei $ ^{9}_{4}Be$ and $ ^{9}_{4}B$ are built mainly on the $ \mathrm{2p_{\frac {3}{2}}} $
shell model orbit\cite{40}. The single - particle orbits 
$\mathrm{1d_{\frac {5}{2}}  }$ and $\mathrm{2s_{\frac {1}{2}}  }$ establish the positive parity states in
$ ^{9}_{4}Be$ and $ ^{9}_{4}B$ isotopes\cite{31}.  In this study, a simplifying
assumption is made that single particle states are built on the 
 $ \mathrm{2p_{\frac {3}{2}}} $ and  $\mathrm{2s_{\frac {1}{2}}  }$. The lattices of algebras in these cases  are obtained by putting j=3/2 and 1/2 in Eq.(5), respectively. 
 
 The Lie algebra corresponding to the $ SU(1,1) $ group is generated by the operators $ S^x $ where $ x = 0 $ and $ \pm1 $. To extend this model, we introduce the $ SU(1,1) $ pairing algebras for s and b bosons as,
 \begin{equation}
 S^{+}(s)=\frac{1}{2}{s^{\dagger}}^2   \quad\quad\quad     S^{-}(s)=\frac{1}{2}s^2       \quad\quad\quad     S^{0}(s)=\frac{1}{2}(s^{\dagger}s+\frac{1}{2})=\frac{1}{2}n_s+ \frac{1}{4}
          \end{equation}
                         
 \begin{equation}
    S^{+}(b)=\frac{1}{2}{b^{\dagger}}.{b^{\dagger}}   \quad\quad\quad     S^{-}(b)=\frac{1}{2}\tilde{b}.\tilde{b}       \quad\quad\quad     S^{0}(b)=\frac{1}{2}(b^{\dagger}\tilde{b}+\frac{3}{2})=\frac{1}{2}n_b+ \frac{3}{4}
             \end{equation} 
   where $ n_s $ and $ n_b $ are the number operators for s and b bosons which satisfy the following commutation relations
                        
     \begin{equation}
        [S^{0},S^{\pm} ]=\pm S^{\pm}  \quad\quad\quad    [ S^{+},S^{-} ]=-2S^{0}          
                 \end{equation}          
  The Casimir operator of $ SU(1,1) $ can be written as
  \begin{equation}
       C_2(SU(1,1))=S^{0}(S^{0}-1)-S^{+}S^{-}      
                   \end{equation}
                   
  The representation is determined by a single number k. Let us assume that $|k\mu\rangle $ is a basis vector of an irrep of $ SU(1,1) $, where κ can be any positive real number, and $ \mu= k,k+1,...$. Then 
    \begin{equation}
         C_2(SU(1,1))|k\mu\rangle=k(k-1)|k\mu\rangle\quad\quad\quad
         S^{0}|k\mu\rangle=\mu|k\mu\rangle
                     \end{equation}                   

 Now, we introduce the operators of infinite dimensional $ SU^{sb} (1,1) $  algebra similar to what has been defined by Pan et al. in ref. \cite{40}, 
  \begin{equation}
        S_n^\pm=c_s^{(2n+1)} S^\pm (s)+c_b^{(2n+1)} S^\pm (b)    
                     \end{equation}
                     
  \begin{equation}
         S_n^0=c_s^{2n} S^0 (s)+c_b^{2n} S^0 (b)    
                      \end{equation}
                      
 where $ c_s $  and $ c_b $ are the real control parameters, and n can be taken to be $ 1, 2, 3, ...  $ .
 To evaluate the energy spectra and transition probabilities, let us consider $ |lw\rangle  $ as the lowest weight state of $ SU^{sb} (1,1) $   algebra which should satisfy
  \begin{equation}
          S^{-}(s)|lw\rangle=0 \quad\quad\quad S^{-}(b)|lw\rangle=0 
                        \end{equation}
                        
 The lowest weight states,$ |lw\rangle_{sb} $ are actually a set of basis vectors of the chain $ U(4)\supset U(3)\supset O(3) \supset O(2) $ which
  \begin{equation}
          |lw\rangle^{B}_{sb}=|N_{B},k_s=\frac{1}{2} (\nu_s+\frac{1}{2}),\mu_s=\frac{1}{2} (n_s+\frac{1}{2}),k_b=\frac{1}{2} (L+\frac{3}{2}),\mu_b=\frac{1}{2} (n_b+\frac{3}{2}) ,LM\rangle
                        \end{equation}
where $ N_{B}=\nu_s+\nu_b $,   $  n_b=L $,      $ n_s=\nu_s=0 $ or $ 1 $. Hence, we have  
  \begin{equation}
           S_{n}^{0}|lw\rangle=(c_s^{(2n)} S^⁰ (s)+c_b^{(2n)} S^⁰(b))|lw\rangle=\Lambda_n^0|lw\rangle
                         \end{equation}
  \begin{equation}
          \Lambda_n^0=(c_s^{(2n)}(\nu_s+\frac{1}{2})+c_b^{(2n)}{(L+\frac{3}{2}))}\frac{1}{2}  
                         \end{equation}

 The following
Hamiltonian for description of negative and positive states in transitional region  is prepared
\begin{equation}
\label{11}
\hat{H }=gS_{0}^{+} S_{0}^{-}+\alpha S_{1}^{0}+\beta\hat{C } _{2}(SO^{B} (3) )+\gamma\hat{C }
_{2}(spin^{BF} (3) )
\end{equation}
For evaluating the eigenvalue of
Hamiltonian Eqs. (17), the eigenstates are considered as
\begin{equation}
\label{14}
|k;\nu_{s}\nu_{b} n_\Delta L JM \rangle= \Theta
S_{x_{1}}^{+}S_{x_{2}}^{+}S_{x_{3}}^{+}...S_{x_{k}}^{+}|lw\rangle^{BF}
\end{equation}
With Clebsch- Gordan (CG) coefficient, we can calculate lowest
weight state, $\mathrm{|lw\rangle^{BF}}$, in terms of boson and fermion
part as

\begin{equation}
\label{18}
|lw\rangle^{BF}=\sum_{j}\sum_{m_{j}=-j}^{m_{j}=+j}C_{m,m-m_{j},m_{j}}^{J,L,j}|lw\rangle_{sb}^{B}|j,m_{j}\rangle
\end{equation}

The $\mathrm{C_{m,m_{L},m_{j}}^{J,L,j} }$symbols represent Clebsch-Gordan coefficients.
 \begin{equation}
          |lw\rangle^{B}_{sb}=|N_{B},k_s=\frac{1}{2} (\nu_s+\frac{1}{2}),\mu_s=\frac{1}{2} (n_s+\frac{1}{2}),k_b=\frac{1}{2} (L+\frac{3}{2}),\mu_b=\frac{1}{2} (n_b+\frac{3}{2}) ,LM\rangle
                        \end{equation}
                        
  \begin{equation}
           S_{n}^{0}|lw\rangle=(c_s^{(2n)} S^⁰ (s)+c_b^{(2n)} S^⁰(b))|lw\rangle=\Lambda_n^0|lw\rangle
                         \end{equation}
  \begin{equation}
          \Lambda_n^0=(c_s^{(2n)}(\nu_s+\frac{1}{2})+c_b^{(2n)}{(L+\frac{3}{2}))}\frac{1}{2}  
                         \end{equation}

The eigenvalues of Hamiltonians Eqs. (17) can then be expressed;

\begin{equation}
\label{24}
E^{(k) }=h^{(k) }+\alpha \Lambda_{1}^{0}+\beta L(L+1)+\gamma J(J+1)
\end{equation}
 where

   \begin{equation}
      h^{(k)}=\sum_{i=1}^{k}\frac{\alpha}{x_i}=\sum_{i=1}^{k}\frac{gc_s^{2}(\nu_s+\frac{1}{2})}{1-c_{s}^{2}x_i }+\frac{gc_b^{2}(L+\frac{3}{2})}{1-c_{b}^{2}x_i }-\sum_{i\neq j}\frac{2g}{x_i-x_j}  \quad\quad g=1
                           \end{equation}

 \subsubsection{The odd-odd nuclei : $ ^{10}_{5}B$}      
  The structure of the odd-odd nuclei may be illustrated as
  an unpaired proton and an unpaired neutron coupled to a
  boson core.
  In this paper, the method described in Refs.\cite{38,39} will be developed and performed to mixed boson-fermion-fermion systems. 
  The approach based on boson-fermion symmetries has
  been also applied to odd-odd systems. On the other, IBFM has been increased to odd-odd nuclei, and mention  to as IBFFM. 
  To simplify computing, the structure of the odd-odd nuclei is described as an unpaired proton and an unpaired neutron coupled to a $ ^{8}_{4}Be$.

  It should be noted that we have investigated the phase transition from rigid to non-rigid shapes in the case that odd proton and
  odd neutron in j=3/2 configurations coupled to 
   core that undergoes a transition from  $ U^{B}(3) $ and $ O^{B}(4) $ condition.
  
  After this, we considered the state that an unpaired proton and an unpaired neutron being in a j=3/2 shell. The algebraic structure underlying our IBFM-1 approach is shown in Eq.(25).The bosons are initially  coupled and so are fermions, and then the compounds of bosons and fermions connect to each other. In Eq.(25) , the chain upper show the state that bosons have $\mathrm{U^{B}(3)}$ dynamical symmetric while bosons in chain lower have $\mathrm{O^{B}(4)}$ dynamical symmetric. 
   \begin{equation}
    U^{B}(4)\otimes( U^{F_{\pi}}(4)\otimes U^{F_{\nu}}(4))\supset\left\{ \begin{array}{l}
  {U^{(B)}}(3)\\
  {O^{(B)}}(4)
  \end{array} \right\}\
  \otimes SU^{F_{\pi\nu}}(4)\supset O^{B}(3)\otimes Sp^{F_{\pi\nu}}(4) \supset O^{B}(3)\otimes SU^{F_{\pi\nu}}(2) \supset  O_{\pi\nu}^{BF}(3)\supset O_{\pi\nu}^{BF}(2) 
   \end{equation} 
 
  The Hamiltonian for the odd-odd nuclei may be
  written as a sum of a boson part and  parts describing
  the residual interaction between boson-fermion and
  fermion-fermion interaction. 
  The 
  Hamiltonian with $j_{\pi}=\frac{3}{2}$ and $j_{\nu}=\frac{3}{2}$ in transitional region  between $\mathrm{U
  (3)-O(4)}$ limits in terms of the casimir operators of  the group chain (Eq.(25)) is prepared
  \begin{equation}
  \label{11}
  \hat{H }=gS_{B,0}^{+} S_{B,0}^{-}+\alpha S_{B,1}^{0}+\beta \hat{C }
  _{2}(SO^{B} (3) )+\rho \hat{C } _{2}(SP^{F_{\pi\nu}} (4))+\delta\hat{C } _{2}(SU^{F_{\pi\nu}} (2))+ \gamma\hat{C } _{2}(O^{BF_{\pi\nu}} (3)) 
  \end{equation}
  
  Eq.(26) is the suggested Hamiltonian for boson -
  fermion-fermion systems  and $\mathrm{\alpha}$
  , $\mathrm{\beta}$ ,$ \mathrm{\delta}$, $\mathrm{\rho}$  and $\mathrm{\gamma}$ are real parameters.
  Hamiltonian Eq.(26) is equivalent to 
  Hamiltonian of  rigid limit when $\mathrm{c_{s}=1}$ and with 
  Hamiltonian of non-rigid limit if $\mathrm{c_{s}=0} $ and $\mathrm{c_{b}}\neq 0 $. So, the $\mathrm{c_{s}\neq
  c_{b} \neq 0}$  situation just corresponds to transitional region.   
   
  For evaluating the eigenvalues of
  Hamiltonian Eqs. (26)  the eigenstates are considered as
  \cite{36,37}
\begin{equation}
\label{13}
|k;\nu_{s},\nu_{b}, (\xi_{1},\xi_{2}), S L,JM \rangle= \Theta
S_{x_{1}}^{+}S_{x_{2}}^{+}S_{x_{3}}^{+}...S_{x_{k}}^{+}|lw\rangle^{BF_{\pi\nu}}
\end{equation}
  $N_{B} , \nu_{b},(\xi_{1},\xi_{2}),S, L, JM $ are quantum numbers of $ U^{B}(4),O^{B}(4),SP^{F_{\pi\nu}}(4), SU^{F_{\pi\nu}} (2), SO^{B} (3)$, $O^{BF_{\pi\nu}}(3)$ and $O^{BF_{\pi\nu}}(2)$, respectively. 
  
 The lowest weight state  $\mathrm{|lw\rangle^{BF_{\pi\nu}}}$ is calculated as:
  \begin{eqnarray}
  \label{17}
  |lw\rangle^{BF_{\pi\nu}}=\sum_{m_{\pi}m_{\nu}m_{\pi\nu}M}{C_{m_{\pi\nu},m_{\pi},m_{\nu}}^{J_{\pi\nu},j_{\pi},j_{\nu}}C_{M,m_{\pi\nu},M_{b}}^{J,J_{\pi\nu},L}|j_{\pi},m_{\pi}\rangle |j_{\nu},m_{\nu}\rangle |lw\rangle_{sb}^{B}}
  \end{eqnarray}

  The eigenvalues of Hamiltonian Eq. (26) can then be expressed;
  
  \begin{equation}
  \label{19}
  E^{(k) }=h^{(k) }+\alpha \Lambda_{1}^{0}+\beta L(L+1)+\rho (\xi_{1\pi\nu}(\xi_{1\pi\nu}+3)+\xi_{2\pi\nu}(\xi_{2\pi\nu}+1))+\delta S(S+1)+\gamma J(J+1))
  \end{equation}

\subsection{The $ SU(1,1) $ expression of Bethe ansatz equations for three-cluster systems}
 
\subsubsection{The odd-A nuclei : $ ^{13}_{6}C$ and $ ^{13}_{7}N$}  
Similar to the covalently bound structures of two $\alpha$ particles and neutrons discussed earlier, we can continue the discussion  with three $\alpha$ particles and neutrons. To start we can regard to the properties of $^{13}C$ for which the experimental evidence has recently been collected \cite{18,31,32}.

 The structure of the linear chain configurations for $^{13}C$ and $^{14}C$ can be based
on the $\alpha+^{9-10}Be$ structure. The carbon isotopes provide an excellent example for testing the concept of full spectroscopy, because for these nuclei, different types of reactions have been studied. The structure of $^{13}C$ will be examined in more detail; it will serve as an image of the identification of cluster states, which are often particle unstable states. By using a large amount of information a complete spectroscopy of the states can be obtained to stimulate the energy of 20MeV. With a dissociation of the single-particle states the ordering of the remaining states into $K=3/2^{-}$ and $3/2^{+}$ bands is obtained.
The mass 13 nuclei is the subject of many shell-model calculations and its sources\cite{17}.

 Generally, normal (negative) parity states in $^{13}C$ (and $^{13}N$) arise from various re-couplings of the nine nucleons in the p-shell. For the positive parity states with a nucleon in the sd-shell, a very satisfactory description is obtained in the weak coupling model, with the lowest two core states $^{12}C$. The basic positive-parity states of $^{13}C$ are obtained by coupling a 2s- or ld-neutron to a core which is either the $0^{+}$ ground or $2^{+}$ first excited state of $^{12}C$ and then antisymmetrizing\cite{17}. These are states of configuration $1s^{4} 1p^{8} 2s$
or $1s^{4} 1p^{8} 1d$; states of configuration $1s^{3} 1p^{10}$ are omitted \cite{18,31,32}.

The negative parity states in the even
- odd nuclei $ ^{13}_{6}C$ and $ ^{13}_{7}N$ are built mainly on the $ \mathrm{1p_{\frac {1}{2}}} $
shell model orbit. The single - particle orbits $ \mathrm{ 1d_{\frac {5}{2}}} $  and
$\mathrm{2s_{\frac {1}{2}}  }$  establish the positive parity states in
odd-mass  $ ^{13}_{6}C$  and  $ ^{13}_{7}N$  isotopes. In this study, a single particle states are built on the 
j=1/2. 
In this case,the boson algebraic structure will be always taken to be $\mathrm{U^{B}(7)}$ coupled to a fermion with angular momentum j=1/2 

Two possible dynamical symmetry limits, $ U^{B}(6)\otimes U^{F}(2) $ and $ O^{B}(7)\otimes U^{F}(2) $, are related to the following two algebraic chains as:
 \begin{equation}
  U^{B}(7)\otimes U^{F}(2)\supset\left\{ \begin{array}{l}
{U^{(B)}}(6)\\
{O^{(B)}}(7)
\end{array} \right\}\
\otimes SU^{F}(2)\supset O^{B}(6)\otimes SU^{F}(2) \supset O^{B}(3)\otimes SU^{F}(2) \supset  Spin^{BF}(3) 
 \end{equation}   

By
employing the generators of Algebra $ \mathrm{\widehat{SU(1,1) }} $ and Casimir operators of subalgebras , the following
Hamiltonian for transitional region between  $ U^{B}(6)\otimes U^{F}(2) $ and $ O^{B}(7)\otimes U^{F}(2) $ is suggested as
\begin{equation}
\label{11}
\hat{H }=gS_{0}^{+} S_{0}^{-}+\alpha S_{1}^{0}+\beta\hat{C } _{2}(O^{B} (6) )+\gamma\hat{C }_{2}(O^{B} (3) )+\rho\hat{C }
_{2}(spin^{BF} (3) )
\end{equation}
The eigenvectors of the Hamiltonian for the excitations can be written as 
\begin{equation}
\label{14}
|k;\nu_{s}\nu_{b_1}\nu_{b_2} n_\Delta L JM \rangle= \mathcal{N}
S_{x_{1}}^{+}S_{x_{2}}^{+}S_{x_{3}}^{+}...S_{x_{k}}^{+}|lw\rangle^{BF}
\end{equation}
where
\begin{equation}
\label{18}
|lw\rangle^{BF}=\pm \sqrt{{\frac {L\pm
m+\frac{1}{2}}{(2L+1)}}}|lw\rangle_{sb_{1}b_{2},m-\frac{1}{2}}^{B}\chi_{+} +
\sqrt{{\frac {L\mp
m+\frac{1}{2}}{(2L+1)}}}|lw\rangle_{sb_{1}b_{2},m+\frac{1}{2}}^{B}\chi_{-}
\end{equation}

These yield the eigenvalues $ E^{(k)}$ of the Hamiltonian Equation (31)  in the form 

  \begin{equation}
    E^{(k)}= h^{(k)}+\alpha\Lambda_1^0 +\beta(\nu(\nu+4))+\gamma L(L+1)+\rho J(J+1)
                          \end{equation}
                          
   where

      \begin{equation}
         h^{(k)}=\sum_{i=1}^{k}\frac{\alpha}{x_i}=\sum_{i=1}^{k}\frac{gc_s^{2}(\nu_s+\frac{1}{2})}{1-c_s^{2}x_i}+\frac{gc_{b_1}^{2}(\nu_{b_1}+\frac{3}{2})}{1-c_{b_1}^{2}x_i}+\frac{gc_{b_2}^{2}(\nu_{b_2}+\frac{3}{2})}{1-c_{b_2}^{2}x_i}-\sum_{i\neq j}\frac{2g}{x_i-x_j} \quad\quad g=1
                              \end{equation}
                              
 \begin{equation}
     \Lambda_1^0={c_s^{2}(\nu_s+\frac{1}{2})+c_{b_1}^{2}(\nu_{b_1}+\frac{3}{2})+c_{b_2}^{2}(\nu_{b_2}+\frac{3}{2})}
                           \end{equation}

\subsubsection{The odd-odd nuclei : $ ^{14}_{7}N$} 

The $^{14}N$ nucleus is considered as a mediator between the cluster nucleus $^{12}C$ and the doubly magic nucleus $^{16}O$ \cite{17,32}. The study of $^{14}N$ nuclei can expand understanding
of the evolution of increasingly complex structures beyond the $ \alpha $-clustering. The
information about the structure of $^{14}N$ has   practical value. As a major component of the Earth’s atmosphere the $^{14}N$ nucleus can be a source of the light rare earth elements Li, Be and B, as well as of deuterium. Production of these elements
happens as a result of bombardment of the atmosphere in its lifetime by high energy
cosmic particles. Hence, the cluster features of the $^{14}N$ fragmentation
can define the affluence of lighter isotopes. Beams of $^{14}N$ nuclei can be used
in radiation therapy, which also gives a applied interest in obtaining detailed data
about the characteristics of the $^{14}N$ fragmentation \cite{18,31}.

Similar to  structures of  $^{10}_{5}B$  discussed before, 
the method outlined in Sec.3 will be  applied for description of $^{14}_{7}N$. The basic positive-parity states of $^{14}N$ are obtained by coupling a $1p_{1/2}$-neutron and $1p_{1/2}$-proton to a  core that undergoes a transition from  $ U^{B}(6) $ and $ O^{B}(7) $ situation.
The lattice of algebras in this case  is shown as
 \begin{equation}
    U^{B}(7)\otimes( U^{F_{\pi}}(2)\otimes U^{F_{\nu}}(2))\supset\left\{ \begin{array}{l}
  {U^{(B)}}(6)\\
  {O^{(B)}}(7)
  \end{array} \right\}\
  \otimes U^{F_{\pi\nu}}(2)\supset O^{B}(3)\otimes SU^{F_{\pi\nu}}(2) \supset O^{B}(3)\otimes SU^{F_{\pi\nu}}(2) \supset  O_{\pi\nu}^{BF}(3)\supset O_{\pi\nu}^{BF}(2) 
   \end{equation}
 The 
  Hamiltonian with $j_{\pi}=\frac{1}{2}$ and $j_{\nu}=\frac{1}{2}$ in transitional region  between $\mathrm{U
  (6)-O(7)}$ limits in terms of the casimir operators of  the group chain (Eq.(55)) is prepared
\begin{equation}
\hat{H}=gS_{0}^{+} S_{0}^{-}+\alpha S_{1}^{0}+\beta\hat{C } _{2}(O^{B} (6) )+\gamma\hat{C }_{2}(O^{B} (3) )+\gamma' \hat{C } _{2}(SU^{F_{\pi\nu}} (6) )+\rho\hat{C }
_{2}(spin^{BF} (3) )
\end{equation}
For evaluating the eigenvalues of
Hamiltonian Eqs. (56)  the eigenstates Eq.(36) are considered with the lowest weight states,$ |lw\rangle_{sb_{1}b_{2}}$.

The eigenvalues of $^{14}_{7}N$ odd-odd three-cluster nuclei can then be obtained as;

\begin{equation}
\label{24}
E^{(k) }=h^{(k) }+\alpha \Lambda_{1}^{0}+\beta \nu(\nu+4)+\gamma L(L+1)+\gamma' S(S+1)+\rho J(J+1)
\end{equation}

\subsection{The $ SU(1,1) $ expression of Bethe ansatz equations for four-cluster systems}
  
\subsubsection{The odd-A nuclei : $ ^{17}_{8}O$ and $ ^{17}_{9}F$}  
In this work, we study the structure $ ^{17}_{8}O $  as Four -$\alpha$ particles and a neutron and  $^{17}_{9}F$ as Four -$\alpha$ particles and a  proton. The negative parity states in the even
- odd nuclei $ ^{17}_{8}O$ and $ ^{17}_{9}F$ are built mainly on the $ \mathrm{1p_{\frac {1}{2}}} $
shell model orbit. The single - particle orbits $ \mathrm{ 1d_{\frac {5}{2}}} $  and
$\mathrm{2s_{\frac {1}{2}}  }$  establish the positive parity states in
odd-mass$ ^{17}_{8}O$ and $ ^{17}_{9}F$isotopes. In this study, a simplifying
assumption is made that single particle states are built on the 
j=1/2. 

In this case,the boson algebraic structure will be always taken to be $\mathrm{U^{B}(10)}$ coupled to a fermion with angular momentum j=1/2 

Two possible dynamical symmetry limits, $ U^{B}(9)\otimes U^{F}(2) $ and $ O^{B}(10)\otimes U^{F}(2) $, are related to the following two algebraic chains as:
 \begin{equation}
  U^{B}(10)\otimes U^{F}(2)\supset\left\{ \begin{array}{l}
{U^{(B)}}(9)\\
{O^{(B)}}(10)
\end{array} \right\}\
\otimes SU^{F}(2)\supset O^{B}(9)\otimes SU^{F}(2) \supset O^{B}(3)\otimes SU^{F}(2) \supset  Spin^{BF}(3) 
 \end{equation}   

In a four-cluster model, the Hamiltonian for the transitional region between $ U^{B}(9)\otimes U^{F}(2) \leftrightarrow O^{B}(10)\otimes U^{F}(2) $ can be considered as 
\begin{equation}
 H=gS_0^+S_0^-+\alpha S_1^0+\beta C_2(O^{B}(9))+\gamma C_2(O^{B}(3))+\rho C_2(Spin^{BF}(3))
\end{equation}
The eigenvalues Equation (41) can be expressed as

   \begin{equation}
     E^{(k)}= h^{(k)}+\alpha\Lambda_1^0 +\beta(\nu(\nu+7))+\gamma L(L+1)+\rho J(J+1)
                           \end{equation} 
                           
\begin{equation}
         h^{(k)}=\sum_{i=1}^{k}\frac{\alpha}{x_i}=\sum_{i=1}^{k}\frac{gc_s^{2}(\nu_s+\frac{1}{2})}{1-c_s^{2}x_i}+\frac{gc_{b_1}^{2}(\nu_{b_1}+\frac{3}{2})}{1-c_{b_1}^{2}x_i}+\frac{gc_{b_2}^{2}(\nu_{b_2}+\frac{3}{2})}{1-c_{b_2}^{2}x_i}
          +\frac{gc_{b_3}^{2}(\nu_{b_3}+\frac{3}{2})}{1-c_{b_3}^{2}x_i}-\sum_{i\neq j}\frac{2g}{x_i-x_j} \quad\quad g=1
                              \end{equation}
                              
 \begin{equation}
     \Lambda_1^0={c_s^{2}(\nu_s+\frac{1}{2})+c_{b_1}^{2}(\nu_{b_1}+\frac{3}{2})+c_{b_2}^{2}(\nu_{b_2}+\frac{3}{2})+c_{b_3}^{2}(\nu_{b_3}+\frac{3}{2})}
                           \end{equation}

\subsection{The fitting procedure}
In order to obtain the numerical results for energy spectra
$\mathrm{(E^{(k) } )}$ of the considered nuclei , a set of non-linear 
Bethe-Ansatz equations (BAE) with k- unknowns for k-pair
excitations must be solved.
 To achieve this aim, we have changed variables in two-cluster nuclei as
$$ \mathrm{C=\frac  {c_{s}}{c_{b}} \leq 1    ,   y_{i}=c_{b}^{2} x_{i}} $$

 In addition, the constants of
Hamiltonian with the least square fitting processes to experimental
data are obtained .
 A useful and simple numerical algorithm for solving the BAE Equations (24) and for extracting of the constants in comparison with the experimental energy spectra of the considered nuclei is based on using of Matlab software which will be outlined simultaneously. To determine the roots of the BAE with the specified values of $ \nu_s $ and $\nu_b $, we solve Equation (24) with the definite values of C and $\alpha$ for $ i=1 $ and then we use the function "syms var" in Matlab to obtain all roots. We then repeat this procedure with different C and $\alpha$ to minimize the root mean square deviation, $ \sigma$, between the calculated energy spectra and the experimental counterparts which explore the quality of the extraction processes. The deviation is defined by the equality   
                                                        \begin{equation}                                                      \sigma=(\frac{1}{N_{tot}} |E_{exp}(i)- E_{cal}(i)|^{2} )^\frac{1}{2}                   \end{equation} 
                                                       $ N_{tot} $ is the number of energy levels which are included in the extraction processes. We have extracted the best set of Hamiltonian's parameters, i.e. $  g $, $\alpha $ and $ \beta $, via the available experimental data.

Similar to the procedures to extract the parameters of the transitional Hamiltonian
in the two-cluster case,Eqs. (35)and (43) were solved
for the i = 1 case with definite values C and $\alpha$ in three and four cluster nuclei.

  \section{Numerical results}
   
   This section presents the results of the numerical solution of the phase transition observable of the algebraic cluster model for the two-, three- and four- body clusters such as level crossing,  expectation values of the boson numbers and Calculated variation behavior of the overlap of the ground-state wave function.  In this research paper, we have taken  $^{9}Be$,$^{9}B$,$^{10}B$ ; $^{13}C$, $^{13}N$, $^{14}N$; $^{17}O$, $^{17}F$ nuclei for the two-, three- and four- cluster.

\subsection{Energy spectrum and level crossing }
   
   In the wake of the theoretical method achieved beforehand, we apply our algebraic model for the cluster model to the $^{9}Be$,$^{9}B$,$^{10}B$ ; $^{13}C$, $^{13}N$, $^{14}N$; $^{17}O$ , $^{17}F$ nuclei.
   
   In our calculation, we have proposed the control parameters $ C $ values in the 0-1 region for the two-, three- and four- body clusters. So, we have analyzed the properties of the $^{9}Be$,$^{9}B$,$^{10}B$ ; $^{13}C$, $^{13}N$, $^{14}N$; $^{17}O$ and $^{17}F$  nuclei in order to investigate the ground- and excited-state spectra related to the models-the best fit which guarantees that the parameters are well determined. Eigenvalues of these models are obtained by solving Bethe Ansatz equations with the extraction processes to experimental data to obtain constants of the Hamiltonian. We explore the best-fitting parameters, which are extracted by the procedures explained in Sect. 3 and the least-square ﬁt to the available experimental data \cite{41,42,43} for the excitation energies for the $^{9}Be$,$^{9}B$,$^{10}B$ ; $^{13}C$, $^{13}N$, $^{14}N$; $^{17}O$ and $^{17}F$  nuclei and the ability of the $ SU(1,1) $-based transitional Hamiltonian  in the reproduction of all considered levels and also the acceptable degree of the extraction procedures.The root mean square deviation, $ \sigma$, between the calculated energy spectra and the experimental counterparts as a function of the control parameter C for these nuclei are shown in Figs (4),(5) and (6). Tables 1-8 show the calculated energy spectra along with the experimental values. Our results show that two-cluster nuclei have  vibrational features but the gamma-unstable rotor character is dominant while a dominancy of dynamical symmetry O(7) exist for three-cluster nuclei, and the four-cluster nuclei have  dominant vibrational features. We see from the figure that in this case the odd particle drives the system toward deformation or sphericity. 
   
   To display how the energy levels change as a function of the control parameter C , the lowest energy levels as a function of C f for the $^{9}Be$,$^{9}B$,$^{10}B$ ; $^{13}C$, $^{13}N$, $^{14}N$, $^{17}O$ and $^{17}F$  nuclei are shown in Figs. 1,2 and 3. The figures show how the energy levels as a function of the control parameter C evolve from one dynamical symmetry limit to the other. It can be seen from the figures that numerous level crossings occur.

\subsection{Expectation values of boson number}
  
 The other quantal order parameters that we mention here are the expectation values of the boson number operators. The expectation values of $ n_b $  are the significant objectives of phase transition. So, we calculated these values to show the treatment of phase transition.
In order to calculate the expectation values of the b-boson number operator, we have to select the suitable roots. Given the proper amount of roots,, we have calculated $ <n_b >$ for two, three and four - clusters in even - even and odd-A nuclei.                 
                                                
Fig.8 shows the expectation values of the b-boson number operator
for the lowest states even-even (left panel) and odd-A nuclei (right
panel) as a function of  control parameter for
$N=10$ bosons.  Fig.9 shows the expectation values of the b-boson number operator
for the lowest states even-even nuclei as a function of  ${C}$ and ${C_{b}}$ control parameters.

The sudden change in these quantities show the phase transition. Figures show that the expectation values of the number of vector-bosons remain approximately constant for a limit and only begin to change rapidly for the other limit. It can be seen
from Fig.8 that in due to the presence of the fermion, the
transition is made sharper for  even-even nuclei while is made smoother for  odd-A nuclei.
We also found
      that the position of the critical point has been shifted by the addition of the odd particle with respect to the even
      case. As a outcome the behavior of the odd and even systems at the corresponding critical points are rather similar.

\subsection{Calculated variation behavior of the overlap of the ground-state wave function} 

 It has been shown previously that the overlap of the ground-state
 wave function with that in the dynamical symmetries
 may also serve as a signature of the phase transition \cite{44,45,46,47}. We have
 calculated the overlap of the ground-state wave functions of
 the Hamiltonians  (17) and (26) in $\left| {\left\langle {{g.s.{C_1}}}
  \mathrel{\left | {\vphantom {{g.s.{C_1}} {g.s.{C_2} = 1}}}
  \right. \kern-\nulldelimiterspace}
  {{g.s.{C_2} }} \right\rangle } \right|$
  with $C_{2}=1$ for  $ ^{9}_{4}Be$ and $ ^{10}_{5}B$. The obtained results
 are illustrated in Fig.10.
It indicates that the largest
absolute value of the derivative of $\left| {\left\langle {{g.s.{C_1}}}
  \mathrel{\left | {\vphantom {{g.s.{C_1}} {g.s.{C_2} = 1}}}
  \right. \kern-\nulldelimiterspace}
  {{g.s.{C_2} }} \right\rangle } \right|$ with
respect to ${C_1}$ occurs around the critical point   ${C_1}=0.6$ for $ ^{9}_{4}Be$ and ${C_1}=0.4$ for $ ^{10}_{5}B$ .

\section{Conclusion}

In this paper, we have studied the phase transitions of the algebraic cluster models. $^{9}Be$,$^{9}B$,$^{10}B$ ; $^{13}C$, $^{13}N$, $^{14}N$; $^{17}O$, $^{17}F$ nuclei were studied in the $ SU(3)\leftrightarrow SO(4) $, $ SU(6)\leftrightarrow SO(7) $ and $ SU(9)\leftrightarrow SO(10) $ phase transitions, related to the description of the relative motion of the cluster configurations. A solvable extended transitional Hamiltonian which is based on $ SU(1,1) $ algebra is proposed to pave the way for of quantum phase transition between the spherical and the deformed phases. The validity of the presented
parameters in the cluster-IBM and cluster-IBFM formulations
has been investigated and it is seen that there
exists a satisfactory agreement between the presented results
and the experimental counterparts. The obtained results
in this study confirm that this ACM technique is
worth extending for investigating  odd-A and odd-odd nuclei. 
So, the  clustering survives the addition of
one and two particles.
we have presented here a  analysis of quantum phase transitions in a system of N bosons and one fermion and shown that  the addition of
 a fermion greatly modifies the critical value at which the phase transition occurs.
Our studies confirm
   the importance of the odd nuclei as necessary signatures to characterize the occurrence of the phase transition and
   to determine the precise position of the critical point. 
         
\textbf {Acknowledgements} 

NA would like to thank Iran National Science Foundation (INSF) for supporting this work financially under the
grant No. 97010553. All authors wish to thank the Research Council of University of Tabriz for the grants.

\clearpage

\begin{table}
\begin{center}
\begin{tabular}{p{15.0cm}}

\footnotesize Table 1. \footnotesize Energy spectra for ${^{9}Be}$. The experimental values are taken form Ref.\cite{43}.\\

\end{tabular}

\begin{tabular}{ccccccccccccc}
\hline
\hline

{$J^{\pi}$ }      &  C=0& C=0.1 & C=0.2 &C=0.3& C=0.4& C=0.5&C=0.6 &C=0.7 &C=0.8&  C=0.9& C=1 &$ E_{exp}  $\\

\hline

\quad\\
$  (3/2)^{-}_{1} $ & 0&0  & 0 & 0 & 0 &  0&0&0&0&0&0&0\ 
\quad\\
$  (5/2)^{-}_{1} $ &1.5889&1.6239&1.7267&1.901&2.1473&2.4758&2.8195&3.2445&3.4093&3.743&4.0505&2.43\
\quad\\
 $  (1/2)^{-}_{1} $& 2.1532&2.1897& 2.2968&2.4665&2.686&2.943& 3.2173 &3.5007& 3.7879& 4.0321& 4.25&2.78\  
\quad\\
$ (3/2)^{-}_{2} $ &6.5591&6.5716 &6.6042 &6.6461 &6.6831 &6.7262 &6.7162 &6.7302 &6.5936 &6.4986 &6.3929 &5.59\
\quad\\
$  (7/2)^{-}_{1} $ & 6.4043&6.3726&6.2991&6.1774&6.0107&5.8307&5.6922&5.4697&6.1993&6.146&6.0962 &6.38	\
\quad\\
 $   (5/2)^{-}_{2} $ & 9.3756 &9.3569& 9.282& 9.1595& 8.9961& 8.825& 8.6027& 8.4211& 8.078& 7.8513& 7.6473&	7.94\
\quad\\
  $  (7/2)^{-}_{2} $ &12.7982& 12.8008 & 12.8056 & 12.8112 & 12.8146 & 12.741 & 12.71 & 12.7323 & 12.5301  &12.4819 & 12.438&11.28\
\quad\\
  $  (1/2)^{-}_{2} $ &16.4439 &16.4738  &16.5639  &16.7055  &16.8882  &17.1396  &17.3836&17.7009 &18.0258 &18.2294  &18.3975&16.97\
\quad\\
 $   (5/2)^{-}_{3} $&17.5033 &17.4853 &17.4308 &17.3415 &17.2226 &17.1377& 16.9803& 16.9338& 16.6338& 16.4517& 16.2801&11.81\
\quad\\
  $ ( 5/2)^{-}_{4}$ &18.3618 &18.3625 & 18.3692 & 18.379 & 18.3902 & 18.3977 & 18.4304 & 18.4193 & 18.6775 & 18.7308 & 18.7788&14.48\
\quad\\
  $  (1/2)^{+}_{1}$ &2.3193 & 2.3899 & 2.5843 & 2.8548 & 3.1364&  3.0205 & 3.1554 & 1.9545  &2.143&  2.3158 & 2.4537&1.68\
\quad\\
 $  (5/2)^{+}_{1}$ &2.0687& 2.1038& 2.2032& 2.3563& 2.5456& 2.7504& 2.941& 3.1169& 3.1547& 3.2556& 3.3277&3.05\
\quad\\
 $  (3/2)^{+}_{1}$ &3.7176& 3.7255& 3.7472& 3.7798& 3.8193& 3.8745& 3.912& 3.9776& 3.9447& 3.9683& 3.9954&4.704\
 \quad\\
 $  (9/2)^{+}_{1}$ &6.2726& 6.2827& 6.3102& 6.3609& 6.4381& 6.5235& 6.6269& 6.7087& 6.6292& 6.7506& 6.8721&6.76\
\quad\\
 $  (5/2)^{+}_{2}$ &15.7784& 15.7847& 15.8014& 15.8179& 15.8223& 15.8264& 15.7848& 15.7465& 15.6724& 15.5554& 15.4296&16.67\
\quad\\
 $  (7/2)^{+}_{1}$ &17.2635& 17.2528& 17.2193& 17.1649& 17.0928& 17.0477& 16.9385& 16.9044& 16.5922& 16.4325& 16.274&17.493\
\quad\\
 $  (9/2)^{+}_{2}$ &20.0563& 20.0434& 20.0079& 19.9559& 19.8958& 19.8227& 19.7702& 19.6737& 19.6714& 19.6731& 19.6896&19.42\
\quad\\
\hline                             
\hline
\end{tabular}
\end{center}
\end{table}

\begin{table}
\begin{center}
\begin{tabular}{p{15.0cm}}

\footnotesize Table 2. \footnotesize Energy spectra for ${^{9}B}$. The experimental values are taken form Ref.\cite{43}.\\

\end{tabular}

\begin{tabular}{ccccccccccccc}
\hline
\hline

{$J^{\pi}$ }      &  C=0& C=0.1 & C=0.2 &C=0.3& C=0.4& C=0.5&C=0.6 &C=0.7 &C=0.8&  C=0.9& C=1 &$ E_{exp}  $\\

\hline

\quad\\
$  (3/2)^{-}_{1} $ & 0&0  & 0 & 0 & 0 &  0&0&0&0&0&0&0\ 
\quad\\
$  (5/2)^{-}_{1} $ &1.1265&1.1913&1.3628&1.6772&1.9321&2.0676&2.1273&2.1008&1.764&1.6621&1.5428&2.36\
\quad\\
 $  (1/2)^{-}_{1} $&2.2418 &2.3055& 2.4924& 2.6239& 3.0569& 3.4863& 3.8993& 4.258 &4.4005 &4.6424 &4.8387&2.75\  
\quad\\
$ (7/2)^{-}_{1} $ &7.3997&7.3858&7.3512&7.4955&7.4425&7.3368&7.3298&7.354&6.8307&6.9002&6.9699 &6.97\
\quad\\
$  (7/2)^{-}_{2} $ & 12.1371&12.1504&12.2106&11.0978&11.3742&11.6108&11.9778&12.3512&11.9667&12.2299&12.4531 &	11.65\
\quad\\
 $   (5/2)^{-}_{2} $ & 12.9455&12.9026& 12.7621 &13.5193 &13.098 &12.5604 &12.098& 11.6862 &10.804 &10.5526 &10.3495&12.19	\
\quad\\
  $  (1/2)^{-}_{2} $ &14.0609 & 14.0776 & 14.1118 & 14.983 & 14.9314 & 14.76 & 14.6158 &14.4548 & 13.8505 & 13.7401  &13.6454&14.01\
\quad\\
  $  (3/2)^{-}_{2} $ &16.076  &16.0852  &16.1148 & 15.4139 & 15.5923  &15.7009&15.8239 & 15.9048 & 15.3796 & 15.4846 & 15.5471&14.66\
\quad\\
 $   (5/2)^{-}_{3} $&13.9636 &13.9107 &13.7407& 14.6736 &14.159& 14.078& 13.576& 13.0789& 16.1277& 15.6881& 15.3283&14.7\
\quad\\
  $ ( 1/2)^{-}_{3}$ &17.5124&  17.5339 & 17.5996&  17.1171&  17.3408&  17.4799&  17.6109&  17.6833&  17.1517&  17.2383& 17.2808&17.076\
 
\quad\\
\hline                             
\hline
\end{tabular}
\end{center}
\end{table}
\clearpage

\begin{table}
\begin{center}
\begin{tabular}{p{15.0cm}}

\footnotesize Table 3. \footnotesize Energy spectra for ${^{10}B}$. The experimental values are taken form Ref.\cite{43}.\\

\end{tabular}

\begin{tabular}{ccccccccccccc}
\hline
\hline

{$J^{\pi}$ }      &  C=0& C=0.1 & C=0.2 &C=0.3& C=0.4& C=0.5&C=0.6 &C=0.7 &C=0.8&  C=0.9& C=1 &$ E_{exp}  $\\

\hline

\quad\\
$  3^{+}_{1} $ & 0&0  & 0 & 0 & 0 &  0&0&0&0&0&0&0\ 
\quad\\
$  1^{+}_{1} $ &0.1583&0.1976&0.3104&0.4811&0.6846&0.892&0.8499&1.0251&1.1804&1.3127&1.4227&0.718\
\quad\\
 $  0^{+}_{1} $& 1.6624& 1.675& 1.7101& 1.7587& 1.809 &1.851& 1.7166& 1.761& 1.7957& 1.8247& 1.852&1.74\  
\quad\\
$  1^{+}_{2} $ &2.0485&2.054&2.0723&2.107&2.1615&2.2366&3.0838&3.1261&3.1794&3.2401&3.3043&2.154\
\quad\\
$  2^{+}_{1} $ & 3.7534&3.7669&3.8042&3.8557&3.9098&3.9572&3.6231&3.7107&3.7825&3.836&3.8713&3.587	\
\quad\\
 $  3^{+}_{2} $ & 5.0882& 5.089& 5.0932& 5.1039& 5.1243& 5.1547& 4.1825& 4.2161&4.2547& 4.2941& 4.3309&4.774	\
\quad\\
  $  2^{+}_{2} $ &5.4188& 5.4261 & 5.4485 & 5.4854 & 5.5349 & 5.5925&  4.5908  &4.6551 & 4.7182 & 4.776  &4.826&5.164\
\quad\\
  $  1^{+}_{3} $ &5.5978 & 5.6099 & 5.6456 & 5.7014  &5.7713  &5.847 & 4.8667  &4.952  &5.0309  &5.0999 & 5.1571&5.18\
\quad\\
 $  2^{+}_{3} $&5.3566& 5.3647 &5.3889& 5.4279& 5.4789& 5.536 &6.1214 &6.1627& 6.193& 6.2107& 6.2167&5.92\
\quad\\
  $  4^{+}_{1} $ &6.427 & 6.4064 & 6.3463 & 6.2538 & 6.1399 & 6.0183 & 6.381 & 6.2598 & 6.1517&  6.0602 & 5.9862&6.025\
\quad\\
 $  3^{+}_{3} $ &6.785&6.7739& 6.7406& 6.6858& 6.6127& 6.5273& 6.9328& 6.8535& 6.7771 &6.7079 &6.6484&7.002\
\quad\\
 $  2^{+}_{4} $ &7.0534& 7.0496& 7.0362& 7.0098& 6.9672& 6.909& 7.3467 &7.2988& 7.2462& 7.1937& 7.1451&7.469\
\quad\\
 $  0^{+}_{2} $ &7.4772& 7.4788& 7.4809& 7.4774& 7.4619& 7.4309& 7.7467 &7.7284& 7.7013& 7.6698& 7.6377&7.56\
\quad\\
\hline                             
\hline
\end{tabular}
\end{center}
\end{table}

\begin{table}
\begin{center}
\begin{tabular}{p{15.0cm}}

\footnotesize Table 4. \footnotesize Energy spectra for ${^{13}C}$. The experimental values are taken form Ref.\cite{44}.\\

\end{tabular}

\begin{tabular}{ccccccccccccc}
\hline
\hline

{$J^{\pi}$ }     &  C=0& C=0.1 & C=0.2 &C=0.3& C=0.4& C=0.5&C=0.6 &C=0.7 &C=0.8&  C=0.9& C=1&$ E_{exp}  $  \\

\hline

\quad\\
 $  (1/2)^{-}_{1} $& 0&0  & 0 & 0 & 0 &  0&0&0&0&0&0&0\ 
\quad\\
 $  (3/2)^{-}_{1} $ & 5.281 &5.3013& 5.3618 &5.4611& 5.597& 5.7661 &5.9635 &6.1827 &6.4158& 6.6536& 6.8868& 3.684507\
\quad\\
  $  (5/2)^{-}_{1} $ &6.2531& 6.2745 &6.3371& 6.4364 & 6.5662 & 6.7186 & 6.8854 & 7.0576 & 7.2263 & 7.3832  &7.5217&7.549\  
\quad\\
 $  (1/2)^{-}_{2} $ & 8.3267& 8.3351& 8.3585& 8.3917&8.4277& 8.4591& 8.4792& 8.4826& 8.466& 8.4278& 8.3693&8.858\
\quad\\
 $  (3/2)^{-}_{2} $&7.7344& 7.7524& 7.8055& 7.891& 8.0048& 8.1418& 8.2965& 8.463& 8.6351 &8.8065 &8.9716&9.899	\
\quad\\
  $  (7/2)^{-}_{1} $ &9.6398& 9.6437 & 9.6558 &9.6771 & 9.7087 & 9.7513 & 9.8046 & 9.8673 & 9.9365  &10.0087 & 10.0799&10.753\
\quad\\
   $  (5/2)^{-}_{2} $ &8.7064 & 8.7255&  8.7807&  8.8663&  8.9739&  9.0943 & 9.2185  &9.3379 & 9.4456 & 9.5363  &9.6066& 6.864\
\quad\\
   $  (1/2)^{-}_{3} $ &11.523& 11.5102& 11.4715&  11.4067 & 11.3158&  11.1993 & 11.0588  &10.8971  &10.7185 & 10.5284 & 10.3334& 12.140\
\quad\\
  $  (3/2)^{-}_{3} $&11.5842&11.5728 &11.5371 &11.4731 &11.3757& 11.2407  &11.0654  &10.8499  &10.5974  &10.3146  &10.0109& 11.748\
\quad\\
  $  (5/2)^{-}_{3} $ &12.5563 & 12.546 & 12.5124 & 12.4484 & 12.3449 & 12.1933  &11.9873&  11.7248  &11.408  &11.0443  &10.6458& 12.130\
\quad\\
  $  (7/2)^{-}_{2} $ &14.5464 & 14.5322 & 14.4922  &14.433  &14.3652  &14.2984 & 14.2425 & 14.2056 & 14.1932  &14.2079  &14.2495& 12.438\
\quad\\
  $  (9/2)^{-}_{1} $ &13.8813 & 13.8717 & 13.8437  &13.8002 & 13.7453 & 13.6838 & 13.6208  &13.5617 & 13.5118  & 13.4756  & 13.4559& 13.410\
\quad\\ 
$  (1/2)^{+}_{1} $ &2.1693&2.2375&2.433&2.7366&3.115&3.5363&3.9701&4.3911&4.7793&5.1207&5.407&3.086\
\quad\\                
$  (5/2)^{+}_{1} $ &4.7845&4.8133&4.897&5.0284&5.1971&5.3914&5.5998&5.8116&6.0174&6.2094&6.3818&3.85\
 \quad\\ 
$  (5/2)^{+}_{2} $ &6.9698&6.9933&7.0612&7.1676&7.3034&7.4589&7.6244&7.7916&7.9534&8.1045&8.2411&6.86\
 \quad\\  
 $  (7/2)^{+}_{1} $ &6.9782& 6.9827& 6.9969 &7.0228& 7.0625& 7.1172 &7.1866 &7.2688 &7.36 &7.4558 &7.5514& 7.492\
\quad\\  
$  (3/2)^{+}_{1} $ &6.8121&6.8379&6.9099&7.013&7.1275&7.2342&7.3175&7.6375&7.3797&7.3543&7.2953&7.677\
 \quad\\
$  (3/2)^{+}_{2} $ &8.9975&8.9956&8.9911&8.9865&8.9855&8.9916&9.0073&9.0331&9.0679&9.1094&9.1547&8.25\
\quad\\
$  (9/2)^{+}_{1} $ &9.7626&9.7462&9.6996&9.6297&9.5459&9.4575&9.3729&9.2986&9.2394&9.1981&9.1757& 9.499.8\
\quad\\
$  (1/2)^{+}_{2} $ &10.0908&11.0819&11.0531&10.9997&10.9167&10.8009&10.6524&10.474&10.2709&10.0503&9.8201 & 10.996\
 \quad\\ 
$  (7/2)^{+}_{2} $ &11.3489&11.3426&11.3253&11.3011&11.2752&11.2523&11.2359&11.2287&11.2319&11.2459&11.2701& 11.848\
\quad\\
$  (5/2)^{+}_{3} $ &12.1396&12.1218&12.0642&11.9574&11.7913&11.559&11.2628&10.9059&10.4998&10.0585&9.5982& 11.950\
\quad\\                                                                                                 
\hline                             
\hline
\end{tabular}
\end{center}
\end{table}

\begin{table}
\begin{center}
\begin{tabular}{p{15.0cm}}

\footnotesize Table 5. \footnotesize Energy spectra for ${^{13}N}$. The experimental values are taken form Ref.\cite{44}.\\

\end{tabular}

\begin{tabular}{ccccccccccccc}
\hline
\hline

{$J^{\pi}$ }      &  C=0& C=0.1 & C=0.2 &C=0.3& C=0.4& C=0.5&C=0.6 &C=0.7 &C=0.8&  C=0.9& C=1 &$ E_{exp}  $\\

\hline

\quad\\
$  (1/2)^{-}_{1} $ & 0&0  & 0 & 0 & 0 &  0&0&0&0&0&0&0\ 
\quad\\
$  (3/2)^{-}_{1} $ &3.91493&3.9237&3.9501&3.9944&4.0569&4.1373&4.2349&4.3484&4.4758&4.6144&4.7612&3.509\
\quad\\
 $  (5/2)^{-}_{1} $& 6.4094& 6.4245& 6.4679& 6.5344& 6.6163& 6.7051& 6.7924& 6.8712& 6.9364& 6.9848& 7.0152&7.387\  
\quad\\
$ (1/2)^{-}_{2} $ &8.8234&8.8198&8.8089&8.7896&8.7609&8.7218&8.6717&8.6102&8.5372&8.4531&8.3585 &8.92\
\quad\\
$  (3/2)^{-}_{2} $ & 8.5307&8.5384&8.5597&8.5895&8.6204&8.6447&8.6555&8.6481&8.6207&8.5736&8.5097 &9.52	\
\quad\\
 $   (5/2)^{-}_{2} $ & 11.0251& 11.0257& 11.02667& 11.0258& 11.0197& 11.0055& 10.9806& 10.9437& 10.8947& 10.8341& 10.7636&10.35	\
\quad\\
  $  (7/2)^{-}_{1} $ &10.6842& 10.674 & 10.6446 & 10.5988 & 10.5409 & 10.4759 & 10.4085 & 10.3431 & 10.2833 & 10.2319 & 10.1905&10.36\
\quad\\
  $  (5/2)^{-}_{3} $ &10.973 & 10.9741 & 10.9762 & 10.9758  &10.9683  &10.949 & 10.9137 & 10.8599 & 10.7865 & 10.694 & 10.5843&9.83\
\quad\\
 $  (3/2)^{-}_{3}$ &12.277& 12.2696& 12.2456& 12.2075& 12.1575& 12.0987& 12.0336& 11.9644& 11.8919& 11.8162& 11.7366&11.87\
\quad\\
 $   (7/2)^{-}_{2} $&12.3174& 12.315& 12.3093& 12.3045& 12.3065& 12.321& 12.3531& 12.406& 12.4807& 12.5766& 12.6911&12.08\
\quad\\
  $ ( 1/2)^{+}_{1}$ &1.9268&  2.0359&  2.3411&  2.783&  2.7814&  3.1872&  3.6265&  4.0786&  4.5232&  4.9407 & 5.3128&2.366\
\quad\\
  $  (5/2)^{+}_{1}$ &4.4871& 4.5078 & 4.5682 & 4.6636 & 4.8281 & 4.9993 & 5.1923 & 5.3997 & 5.6129 & 5.8223 & 6.0182&3.55\
\quad\\
 $  (5/2)^{+}_{2}$ &6.5506& 6.5508& 6.5548& 6.5697& 6.5916& 6.6313& 6.6888& 6.7646& 6.8581& 6.9677& 7.0908&6.898\
\quad\\
 $  (3/2)^{+}_{1}$ &5.6715 &5.7175 &5.8489 &6.0477 &6.2646 &6.5261 &6.7931 &7.0498 &7.284 &7.4869 &7.6524&7.166\
 \quad\\
 $  (7/2)^{+}_{1}$ &6.5819& 6.5832& 6.5885& 6.601& 6.6326& 6.6722& 6.7268& 6.7964& 6.879& 6.9711& 7.00678&7.9\
\quad\\
 $  (3/2)^{+}_{2}$ &7.735& 7.7415& 7.7645& 7.8111& 7.8095& 7.8802& 7.9833& 8.1206& 8.2919& 8.495& 8.725 & 7.9\
\quad\\
 $  (9/2)^{+}_{1}$ &9.2198& 9.2066& 9.1702& 9.1182& 9.0183& 8.9301& 8.8399& 8.7539& 8.6783& 8.6193& 8.5822& 9\
\quad\\
 $  (1/2)^{+}_{2}$ &8.8095& 8.8399& 8.9211& 9.0282& 9.1903& 9.2946& 9.3506& 9.344& 9.2709& 9.1323& 8.9362& 10.25\
\quad\\
 $  (5/2)^{+}_{2}$ &11.1985& 11.184& 11.1341& 11.0355& 11.0043& 10.8515& 10.6411& 10.3714& 10.0437& 9.6625& 9.2364&6.382\
\quad\\
 $  (3/2)^{+}_{3}$ &12.383& 12.3747& 12.3437& 12.277& 12.22& 12.1004& 11.9356& 11.7274& 11.4775& 11.1898& 10.8706&13.962\
\quad\\
 $  (1/2)^{+}_{3}$ &12.9241& 12.8963& 12.8152& 12.6887& 12.552& 12.3909& 12.2285& 12.0733& 11.9298& 11.7993& 11.6816& 11.86\
\quad\\
\hline                             
\hline
\end{tabular}
\end{center}
\end{table}

\begin{table}
\begin{center}
\begin{tabular}{p{15.0cm}}

\footnotesize Table 6. \footnotesize Energy spectra for ${^{14}N}$. The experimental values are taken form Ref.\cite{44}.\\

\end{tabular}

\begin{tabular}{ccccccccccccc}
\hline
\hline

{$J^{\pi}$ }     &  C=0& C=0.1 & C=0.2 &C=0.3& C=0.4& C=0.5&C=0.6 &C=0.7 &C=0.8&  C=0.9& C=1&$ E_{exp}  $  \\

\hline

\quad\\
$  1^{+}_{1} $ & 0&0  & 0 & 0 & 0 &  0&0&0&0&0&0&0\ 
\quad\\
$  0^{+}_{1} $ &0.9068&0.9437&1.0498&1.2111&1.4077&1.6176&1.8206&2.0017&2.1516&2.2668&2.3477&2.31\
\quad\\
 $  1^{+}_{2} $ &4.0411& 4.0567& 4.1026& 4.1759& 4.2725& 4.3869& 4.5136& 4.6472 &4.7834 &4.9184 &5.0494&3.945\  
\quad\\
$  1^{+}_{3} $ & 6.1934&6.2082&6.2533&6.3309&6.4422&6.5865&6.7598&6.9554&7.1654&7.3814&7.5963&6.198\
\quad\\
$  3^{+}_{1} $ &6.8276&6.8356&6.8592&6.8969&6.9469&7.0079&7.0783&7.1567&7.2418&7.3316&7.4244&6.444	\
\quad\\
 $  2^{+}_{1} $ &7.5001&7.4986 &7.4935 &7.4837 &7.4681 &7.4469 &7.4214 &7.3939 &7.3666 &7.3417 &7.3209&7.028\
\quad\\
  $  0^{+}_{2} $ &8.4072 & 8.4102 & 8.4181 & 8.4285 & 8.4386 & 8.4461 & 8.4498 & 8.4498 & 8.4466 & 8.4412 & 8.4343&8.617\
\quad\\
  $  5^{+}_{1} $ &8.9856& 9.0064 & 9.069 & 9.1735 & 9.3183 & 9.4993 & 9.7093 & 9.9395 & 10.1801 & 10.4225 & 10.659&8.963\
\quad\\
 $  2^{+}_{2} $&9.6524& 9.6565& 9.6683& 9.6868& 9.7102& 9.7367& 9.7644& 9.792& 9.8187& 9.8439& 9.8678&8.979\
\quad\\
  $  3^{+}_{2} $ &8.9799 & 8.9936 & 9.034 & 9.1 & 9.1891 & 9.2977 & 9.4213 & 9.5549 & 9.6939 & 9.8338 & 9.9713&11.516\
\quad\\
  $ 2^{+}_{3} $ &8.7507 & 8.7566 & 8.7724 & 8.7933 & 8.8135 & 8.8285 & 8.8359 & 8.8358 & 8.8295 & 8.8187 & 8.8048&9.172\
\quad\\
  $  0^{-}_{1} $ &4.8807 & 4.8818 & 4.8848 & 4.889 & 4.8931 & 4.5552 & 4.5833 & 4.5391 & 4.4793 & 4.4083 & 4.3309&4.913\
\quad\\ 
 $  2^{-}_{1} $ &5.2409& 5.2487& 5.2701& 5.2989& 5.3268& 5.1577& 5.1426& 5.1206& 5.082& 5.0307& 4.9712&5.106\
\quad\\
 $  1^{-}_{1} $ &5.3011& 5.3067& 5.3241& 5.3541& 5.397& 6.1234& 6.217& 6.3487& 6.4619& 6.55& 6.6119&5.691\
\quad\\
 $  3^{-}_{1} $ &5.3586& 5.3439& 5.2996& 5.2254& 5.1215& 5.5848& 5.4038& 5.1965 &4.9694 &4.7346 &4.5037&5.833\
\quad\\ 
$  1^{-}_{2} $ &7.923&7.9096&7.8694&7.8031&7.712&7.637&7.4907&7.3329&7.161&6.9843&6.8111&8.061\
\quad\\              
$  4^{-}_{1} $ &8.4881&8.4902&8.4974&8.5117&8.5354&8.5925&8.6625&8.7457&8.8354&8.9265&9.0142&8.489\
\quad\\ 
$  0^{-}_{2} $ &8.8404&8.8376&8.8292&8.8158&8.7982&8.6968&8.6502&8.6098&8.5674&8.5247&8.4838&8.8\
\quad\\ 
$  3^{-}_{2} $ &9.0687&9.0593&9.0315&8.9868&8.9276&8.4288&8.3516&8.2847&8.2087&8.1278&8.046&8.907\
\quad\\ 
$  2^{-}_{2} $ &8.951&8.9553&8.9678&8.987&9.011&9.7652&9.8113&9.8435&9.8715&9.8939&9.9107&9.129\
\quad\\ 
$  2^{-}_{3} $ &9.6379&9.6322&9.6154&9.5887&9.5534&9.3507&9.2574&9.1766&9.0917&9.0065&8.9246&9.388\
\quad\\         
\hline                             
\hline
\end{tabular}
\end{center}
\end{table}

\begin{table}
\begin{center}
\begin{tabular}{p{15.0cm}}

\footnotesize Table 7. \footnotesize Energy spectra for ${^{17}O}$. The experimental values are taken form Ref.\cite{45}.\\

\end{tabular}

\begin{tabular}{ccccccccccccc}
\hline
\hline

{$J^{\pi}$ }      &  C=0& C=0.1 & C=0.2 &C=0.3& C=0.4& C=0.5&C=0.6 &C=0.7 &C=0.8&  C=0.9& C=1 &$ E_{exp}  $\\

\hline

\quad\\
\ 
\quad\\
$  (1/2)^{-}_{1} $ &3.0634&3.0633&3.0628&3.0618&3.0598&3.0565&3.0518&3.0459&3.0392&3.032&3.025& 3.05536\
\quad\\
 $  (5/2)^{-}_{1} $& 3.6874& 3.685& 3.6778& 3.6663& 3.6513& 3.6335& 3.6141& 3.5941& 3.5747& 3.5566& 3.5407& 3.8428\  
\quad\\
$ (3/2)^{-}_{1} $ &4.4415&4.4473&4.464&4.4898&4.522&4.5576&4.5935&4.6271&4.6565&4.6805&4.6989 & 4.5538\
\quad\\
 $   (9/2)^{-}_{1} $ & 5.3837& 5.3882& 5.4015& 5.424& 5.4555& 5.4957& 5.5437& 5.5981& 5.6566& 5.7173& 5.777& 5.2158	\
\quad\\
  $  (3/2)^{-}_{2} $ &5.4204 & 5.42 & 5.4184 & 5.4147 & 5.4079 & 5.3969 & 5.3811& 5.3606 & 5.336& 5.3085& 5.2796& 5.3792\
\quad\\
  $  (7/2)^{-}_{2} $ &6.2395& 6.234 & 6.2176 & 6.1906 & 6.1538 & 6.1083 & 6.0556&5.997 & 5.9364& 5.8742  &5.8129& 5.697\
\quad\\
 $  (5/2)^{-}_{2}$ &5.4857& 5.483& 5.4756& 5.4634& 5.4469& 5.4265& 5.4028& 5.3766& 5.3487& 5.32& 5.2913& 5.7328\
\quad\\
 $   (1/2)^{-}_{2} $&5.9402& 5.9514& 5.9846& 6.0384& 6.1103& 6.1967& 6.2931& 6.3944& 6.4959& 6.5939& 6.6851& 5.939\
\quad\\
  $ ( 7/2)^{-}_{2}$ &6.7411 & 6.7401 & 6.7372& 6.7329 & 6.7279 & 6.7229 & 6.1787&  6.7158&  6.7144&  6.7147&  6.7126& 6.972\
\quad\\
  $  (5/2)^{-}_{3}$ &7.2834 & 7.2764& 7.2553 & 7.2206& 7.1729& 7.1137& 7.0448& 6.9687 & 6.888 & 6.8055 & 6.7236& 7.1657\
\quad\\
 $  (3/2)^{+}_{1}$ &5.2822& 5.2825& 5.2837& 5.286& 5.29& 5.2961& 5.3051& 5.3172& 5.3327& 5.3516& 5.3733& 5.0848\
\quad\\
 $  (3/2)^{+}_{2}$ &6.1557& 6.1588& 6.1681& 6.1831& 6.2034& 6.228& 6.256& 6.2863& 6.3177& 6.3491& 6.3792& 5.8691\
 \quad\\
 $  (1/2)^{+}_{2}$ &6.4896& 6.4885& 6.4852& 6.4797& 6.472& 6.4621& 6.4501& 6.4365& 6.4217& 6.4062& 6.3909& 6.356\
\quad\\
 $  (5/2)^{+}_{2}$ &6.4726& 6.4722& 6.471& 6.4685& 6.4645& 6.4585 &6.4501& 6.439& 6.4254& 6.4093& 6.3915& 6.862\
\quad\\
 $  (3/2)^{+}_{3}$ &7.0292& 7.0289& 7.028& 7.0267& 7.0251& 7.0235& 7.0221& 7.0209& 7.0201& 7.0197& 7.0197& 7.202\
\quad\\
 $  (5/2)^{+}_{3}$ &7.3461& 7.3466& 7.3479& 7.3503& 7.3538& 7.3585& 7.3645& 7.3717& 7.3797& 7.3884& 7.3972& 7.3792\
\quad\\
 $  (7/2)^{+}_{1}$ &7.4405& 7.4388& 7.4339& 7.4261& 7.4159& 7.404& 7.3909& 7.3775& 7.3644& 7.352& 7.3408& 7.576\
\quad\\
 $  (1/2)^{+}_{3}$ &8.2366& 8.2366& 8.2366& 8.2365& 8.2361& 8.2353& 8.2339& 8.2317& 8.2286& 8.2246& 8.2198& 7.956\
\quad\\
\hline                             
\hline
\end{tabular}
\end{center}
\end{table}

\begin{table}
\begin{center}
\begin{tabular}{p{15.0cm}}

\footnotesize Table 8. \footnotesize Energy spectra for ${^{17}F}$. The experimental values are taken form Ref.\cite{45}.\\

\end{tabular}

\begin{tabular}{ccccccccccccc}
\hline
\hline

{$J^{\pi}$ }      &  C=0& C=0.1 & C=0.2 &C=0.3& C=0.4& C=0.5&C=0.6 &C=0.7 &C=0.8&  C=0.9& C=1 &$ E_{exp}  $\\

\hline

\quad\\
$  (3/2)^{+}_{1} $ & 4.6461&4.6439&4.6372&4.6264&4.612&4.5946&4.5751&4.5543&4.5332&4.5126&4.4934& 5\ 
\quad\\
$  (3/2)^{+}_{2} $ &5.8957&5.8973&5.9019&5.9098&5.9209&5.9354&5.9531&5.9734&5.996&6.0199&6.0443& 5.820\
\quad\\
 $  (1/2)^{+}_{1} $& 6.4882& 6.4882& 6.4881& 6.4874& 6.4854& 6.4813& 6.4745& 6.4648& 6.4521& 6.4367& 6.4194 & 6.56\  
\quad\\
$ (5/2)^{+}_{1} $ &6.1578&6.1567&6.1535&6.1484&6.1419&6.1344&6.1265&6.1185&6.1107&6.1033&6.0964 & 6.697\
\quad\\
$  (3/2)^{+}_{3} $ & 7.1453&7.1448&7.1432&7.1408&7.1378&7.1343&7.1308&7.1273&7.1241&7.1212&7.1186 & 6.774	\
\quad\\
 $   (3/2)^{+}_{4} $ &7.4112& 7.4115& 7.4122& 7.4134& 7.415& 7.417& 7.4192& 7.4217& 7.4242& 7.4268& 7.4293&	7.356\
\quad\\
  $  (3/2)^{+}_{5} $ &8.0037&  8.0043&  8.0061&  8.0089&  8.0126&  8.017&  8.0218&  8.027&  8.0323&  8.0375&  8.0426& 7.479\
\quad\\
  $  (1/2)^{-}_{1} $ &3.037&  3.0378&  3.0401&  3.0439&  3.0495&  3.0566&  3.0653&  3.0752&  3.0861&  3.0976&  3.1084& 3.104\
\quad\\
 $  (5/2)^{-}_{1}$ &3.5124& 3.5192& 3.5395& 3.5729& 3.618& 3.6726& 3.7335& 3.797& 3.8595& 3.9181& 3.966& 3.857\
\quad\\
 $   (3/2)^{-}_{1} $&4.773& 4.7712& 4.7658& 4.7567& 4.7439& 4.7274& 4.7077& 4.6856& 4.6619& 4.6378& 4.6165& 4.64\
\quad\\
  $ ( 9/2)^{-}_{1}$ &5.2934&  5.2927&  5.2906&  5.2873&  5.283&  5.2782&  5.2736&  5.2697&  5.2671&  5.266&  5.2663& 5.22\
\quad\\
  $  (3/2)^{-}_{2}$ &5.6038 & 5.6018& 5.5956 & 5.5851 & 5.5698 & 5.5498 & 5.5252 & 5.4968 & 5.4658 & 5.4337 & 5.405& 5.488\
\quad\\
 $  (7/2)^{-}_{1}$ &5.9012& 5.8987& 5.8911& 5.8784& 5.8607& 5.8382& 5.8119& 5.7828 &5.7524& 5.7219& 5.6955& 5.672\
\quad\\
 $  (5/2)^{-}_{2}$ &5.7277& 5.7255& 5.7187& 5.7073& 5.691& 5.67& 5.6446& 5.616& 5.5852& 5.5538& 5.526& 5.682\
 \quad\\
 $  (1/2)^{-}_{2}$ &5.97& 5.9718& 5.9771& 5.9857& 5.9974& 6.0115& 6.0275& 6.0445& 6.0617& 6.0785& 6.0929& 6.037\
\quad\\
 $  (5/2)^{-}_{3}$ &7.1098& 7.111& 7.1148& 7.1212& 7.1302& 7.1419& 7.1559& 7.1714& 7.1876& 7.2038& 7.2176& 7.027\
\quad\\
 $  (7/2)^{-}_{1}$ &7.2833& 7.2842& 7.2871& 7.2923& 7.2999& 7.3102& 7.3231& 7.3382& 7.3548& 7.3719& 7.3871& 7.546\
\quad\\
\hline                             
\hline
\end{tabular}
\end{center}
\end{table}

\begin{figure}
\begin{center}
\includegraphics[height=6cm]{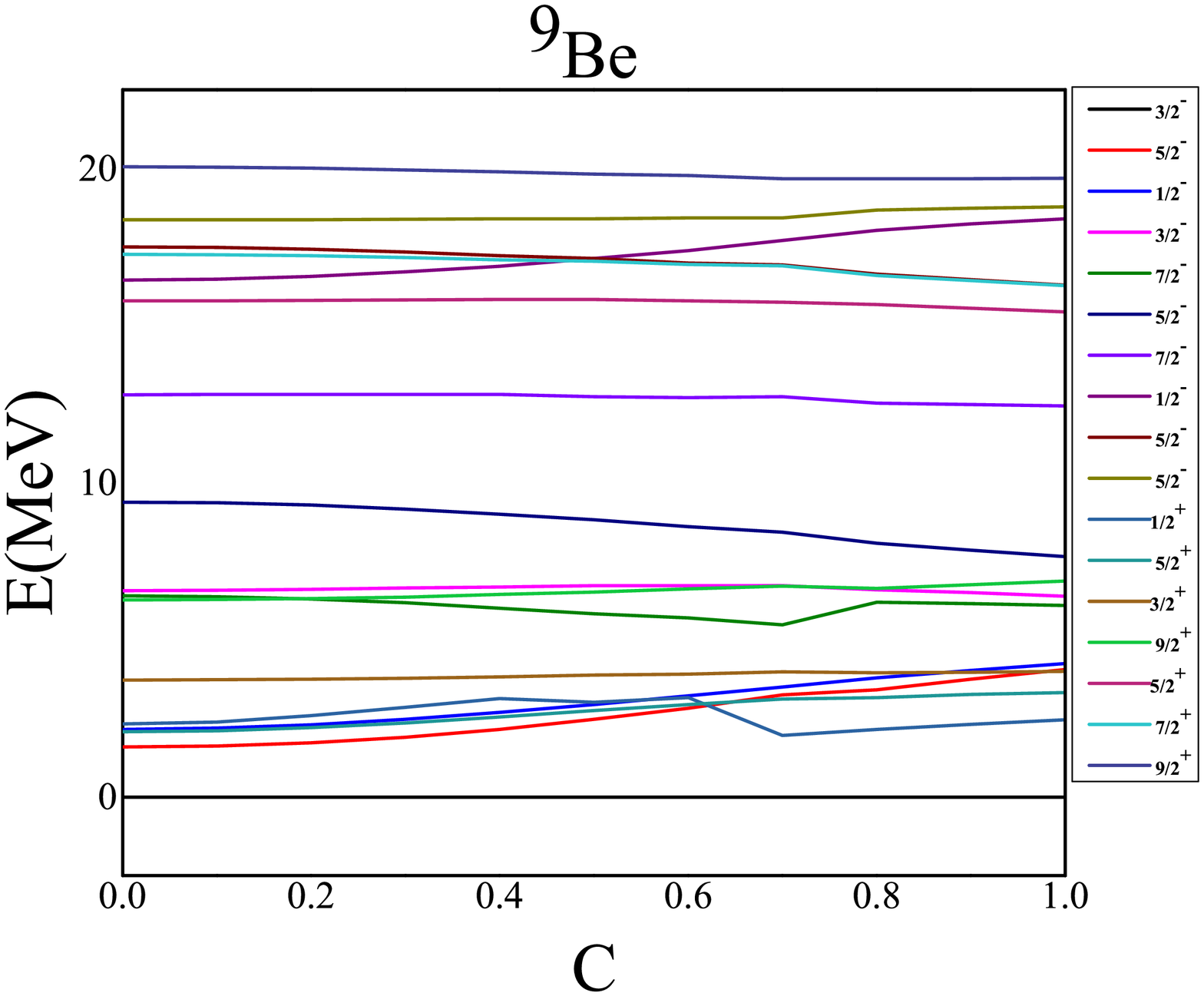}
\includegraphics[height=6cm]{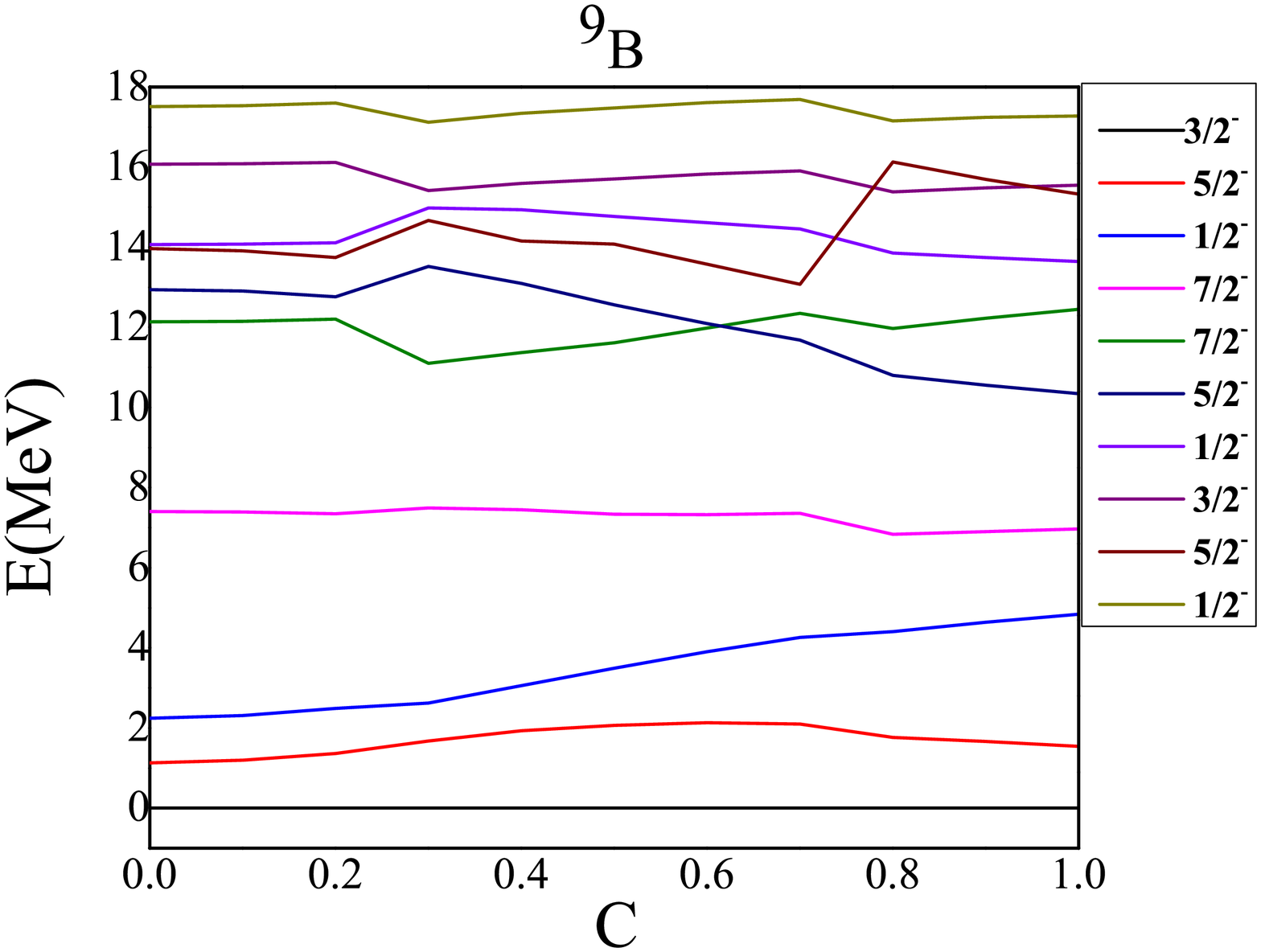}
\includegraphics[height=6cm]{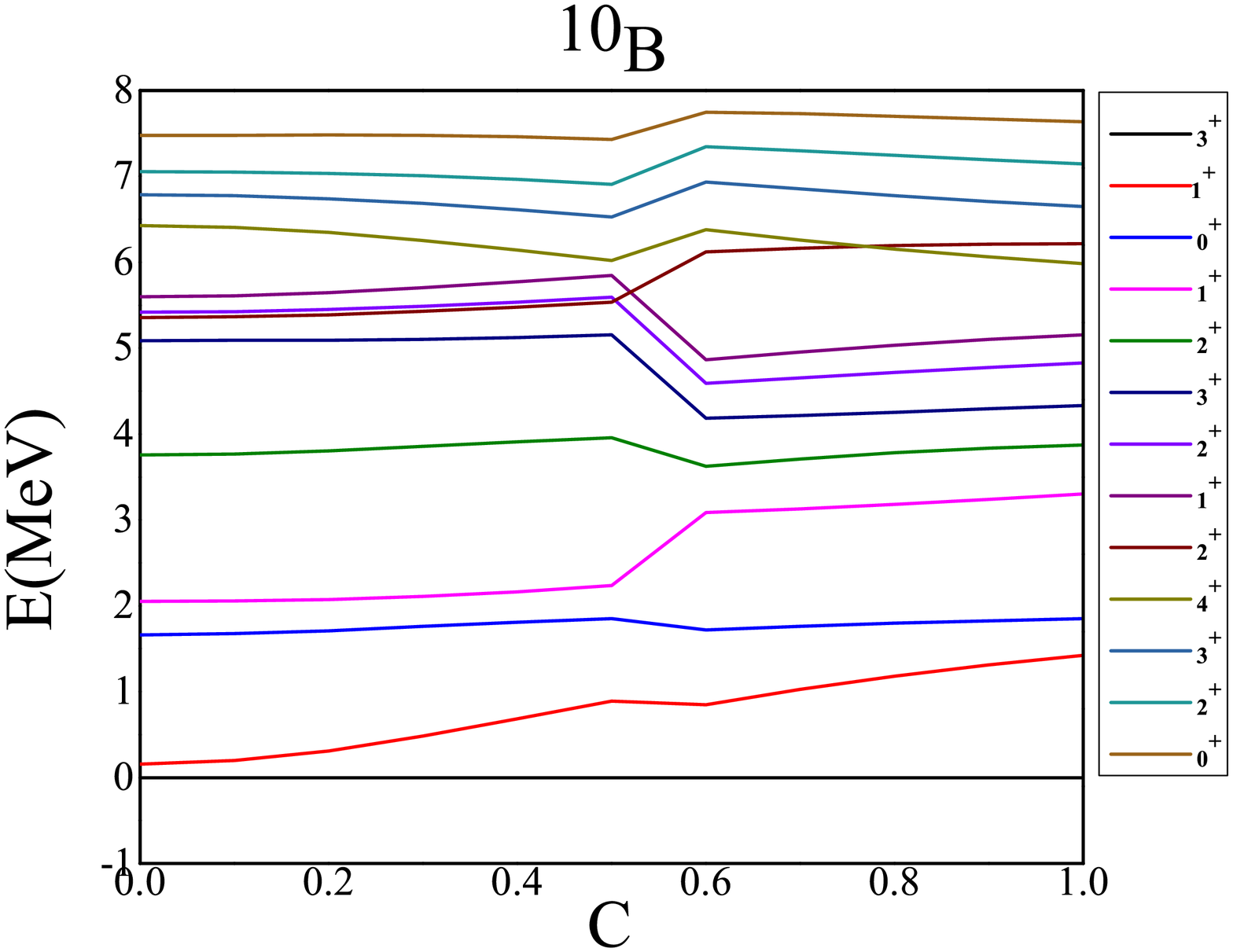}
\caption{Energy levels as a function of the control parameter C for two-cluster nuclei.}\label{fig:1}
\end{center}
\end{figure}

\begin{figure}
\begin{center}
\includegraphics[height=6cm]{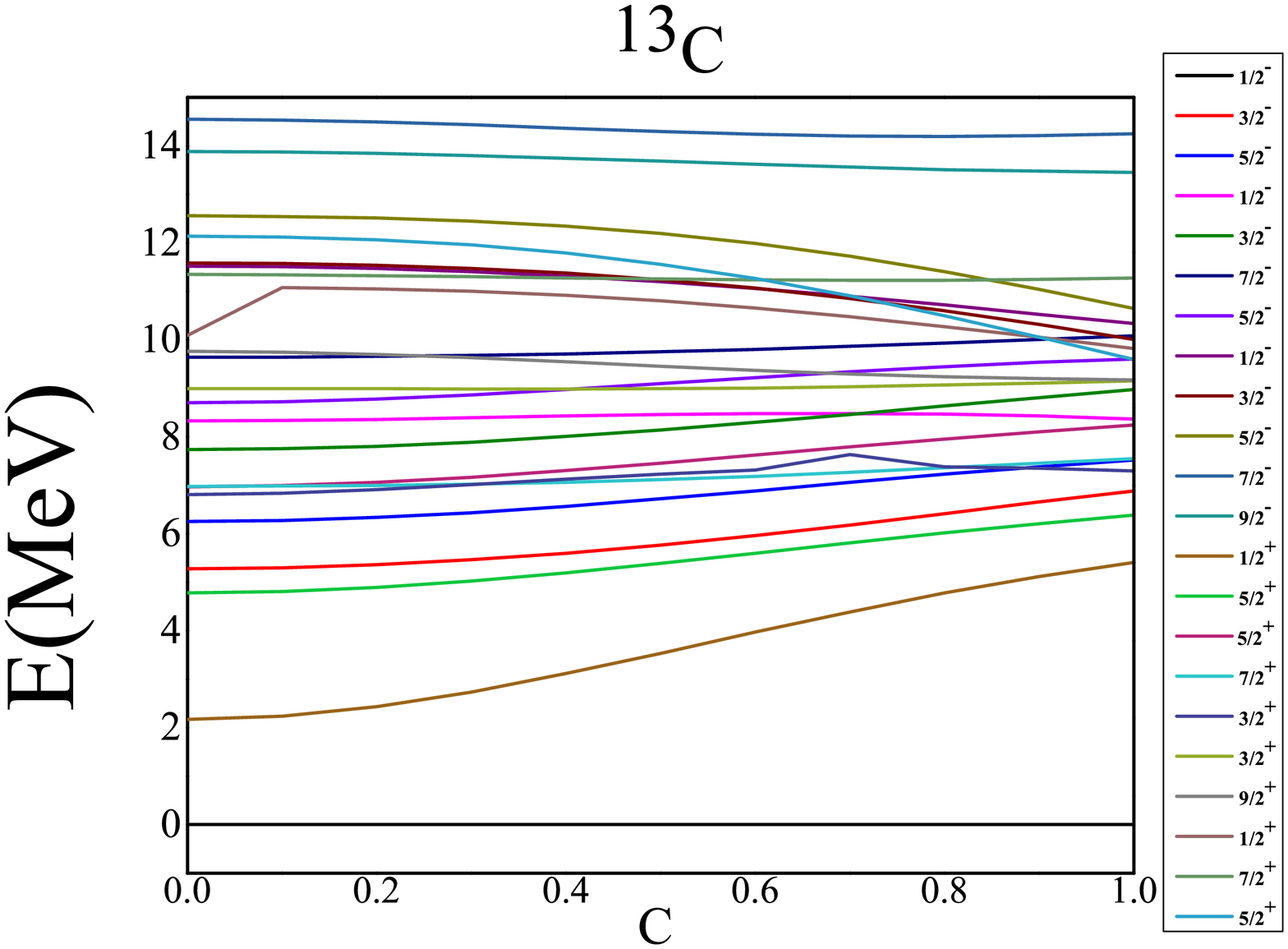}
\includegraphics[height=6cm]{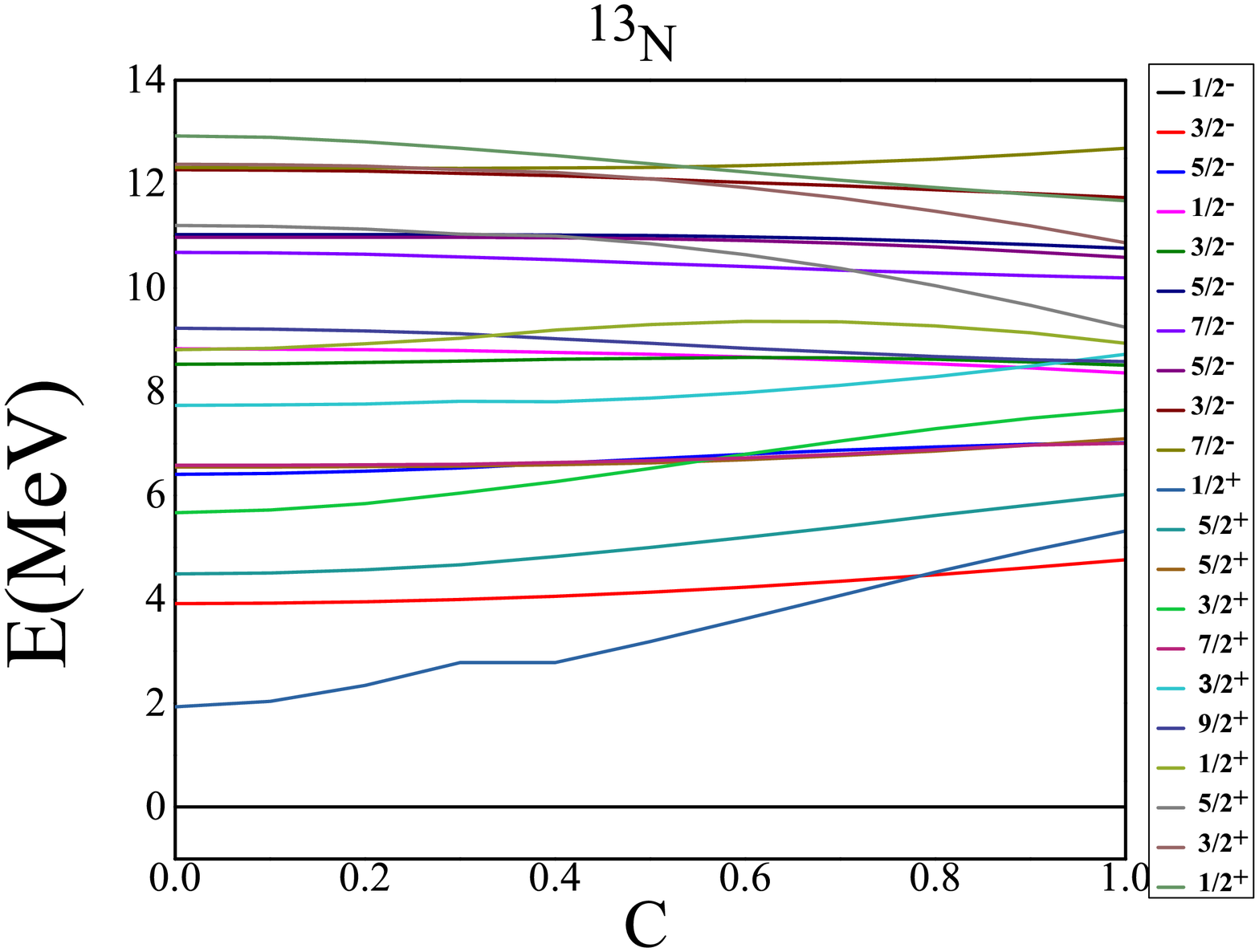}
\includegraphics[height=6cm]{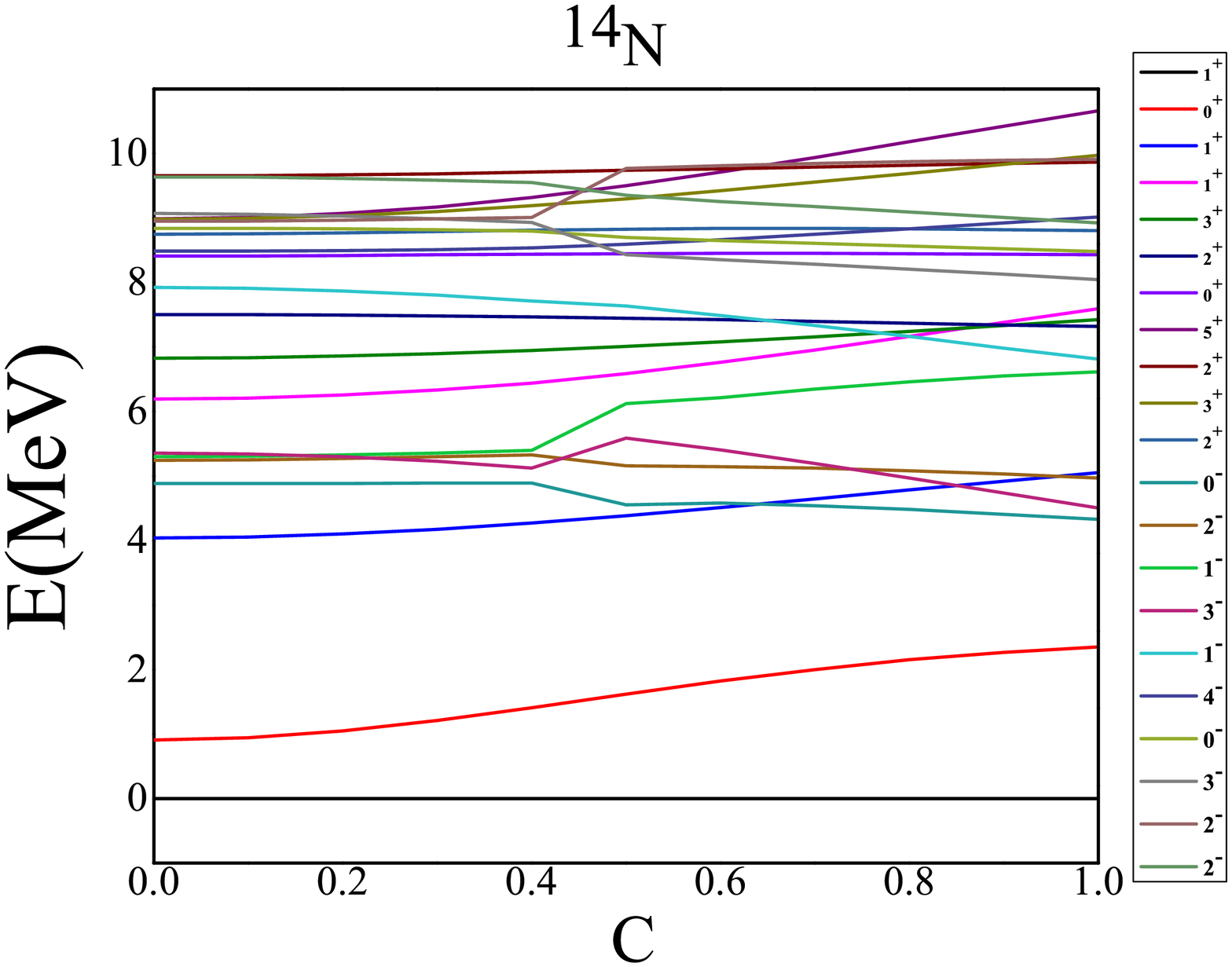}
\caption{Energy levels as a function of the control parameter C  for three-cluster nuclei.}\label{fig:2}
\end{center}
\end{figure}

\begin{figure}
\begin{center}
\includegraphics[height=6cm]{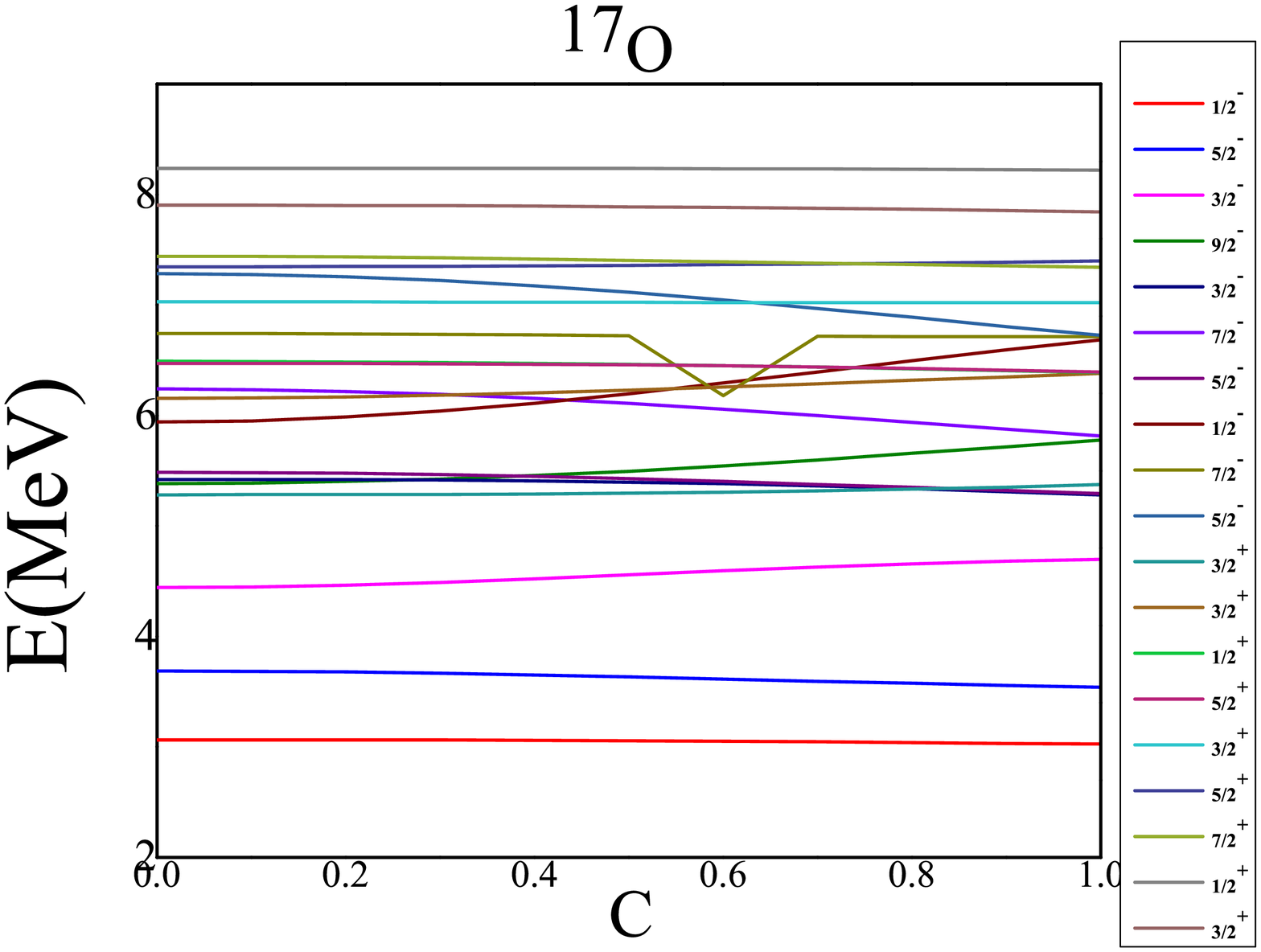}
\includegraphics[height=6cm]{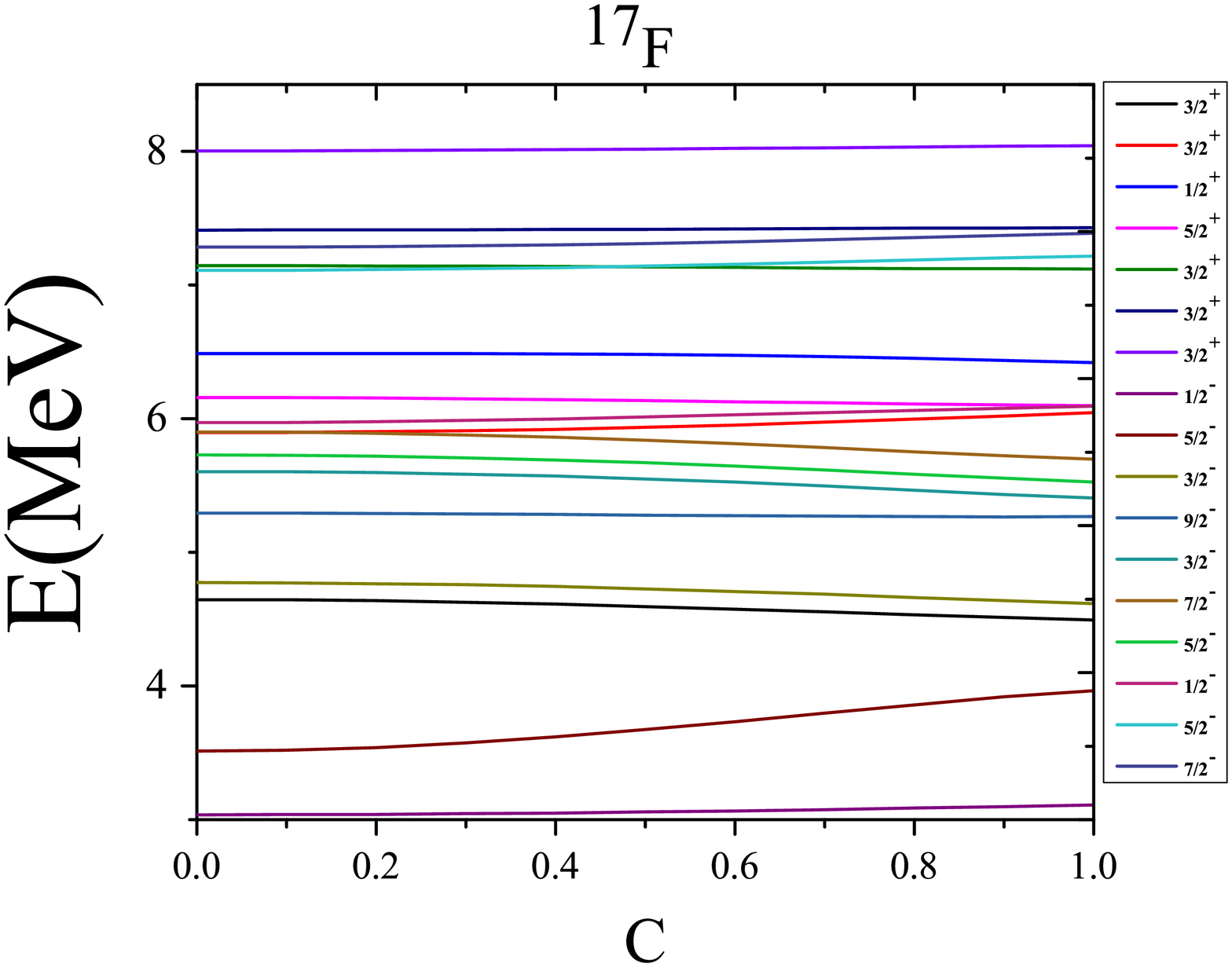}
\caption{Energy levels as a function of the control parameter C for four-cluster nuclei .}\label{fig:3}
\end{center}
\end{figure}

\begin{flushleft}
\end{flushleft}

\begin{figure}
\begin{center}
\includegraphics[height=6cm]{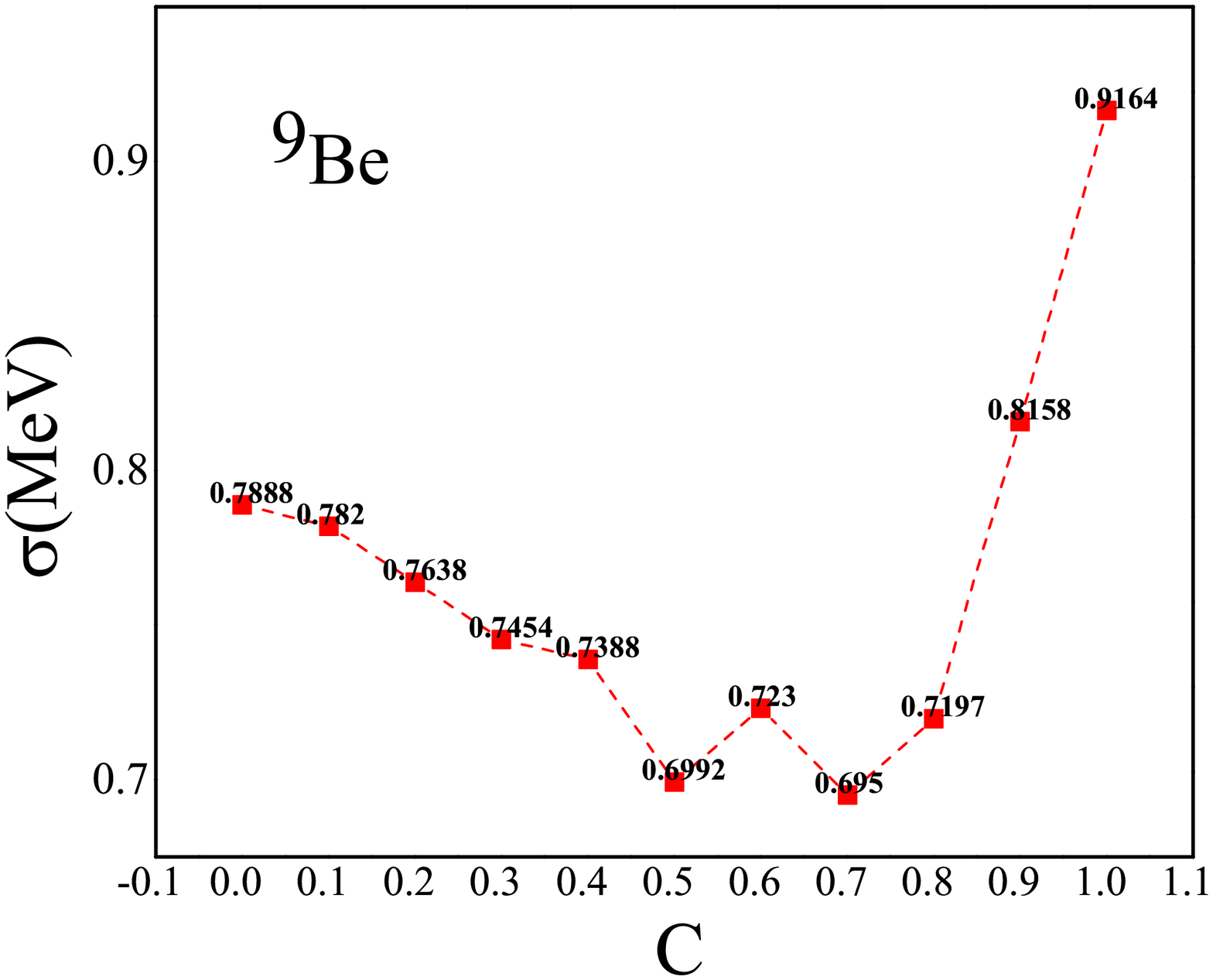}
\includegraphics[height=6cm]{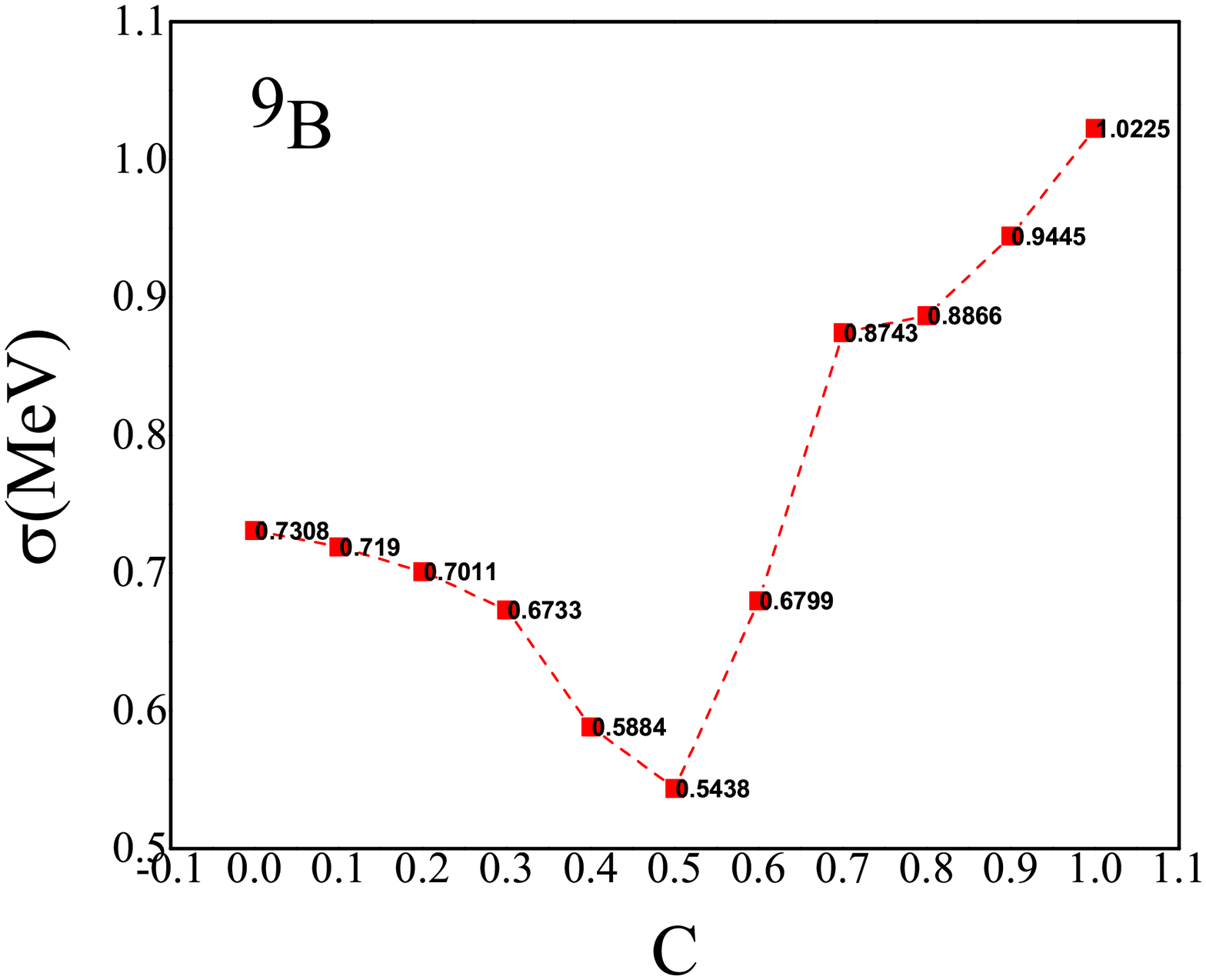}
\includegraphics[height=6cm]{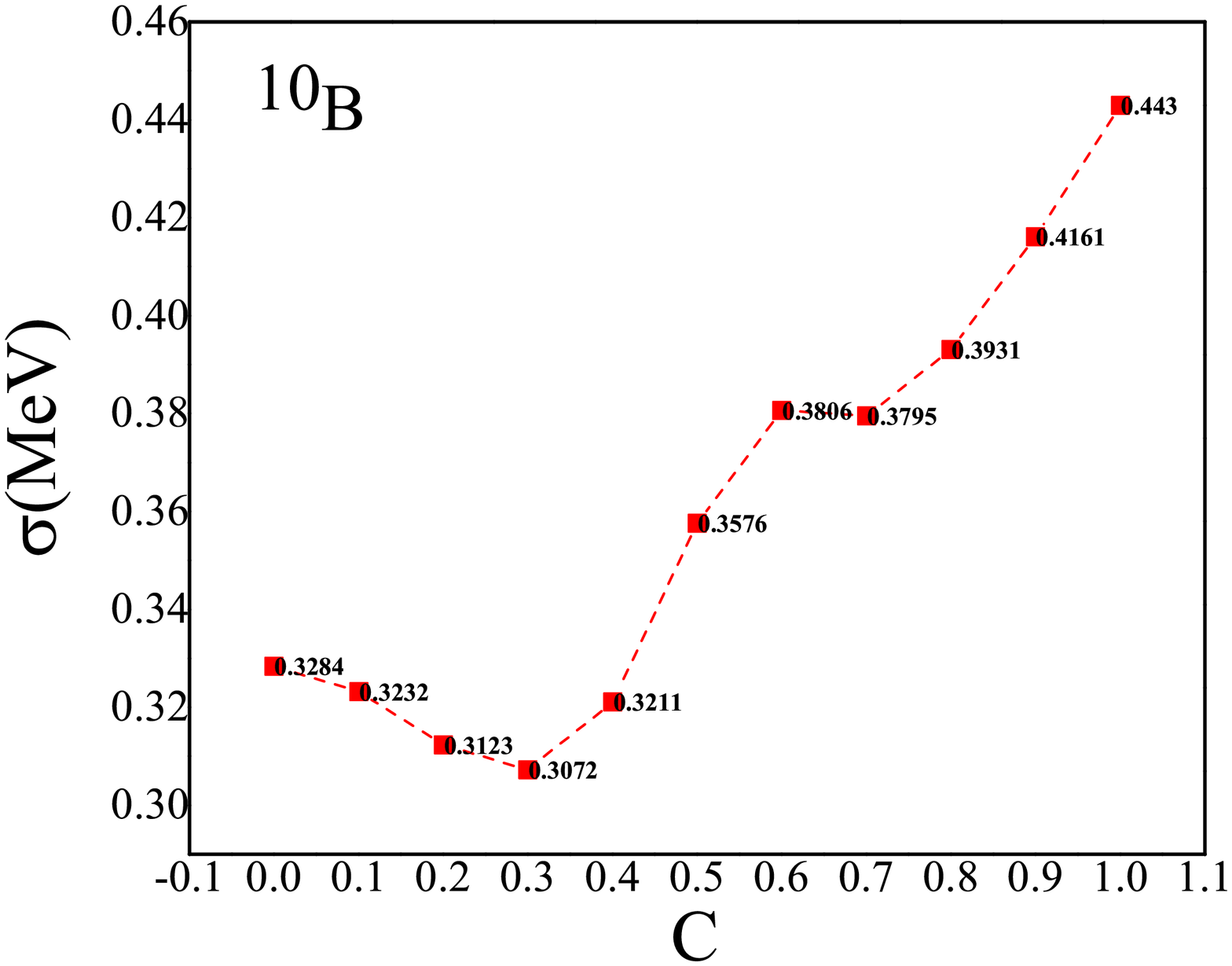}
\caption{ The root mean square deviation as a function of the control parameter C for two-cluster nuclei.}\label{fig:4}
\end{center}
\end{figure}

\begin{flushleft}
\end{flushleft}

\begin{figure}
\begin{center}
\includegraphics[height=6cm]{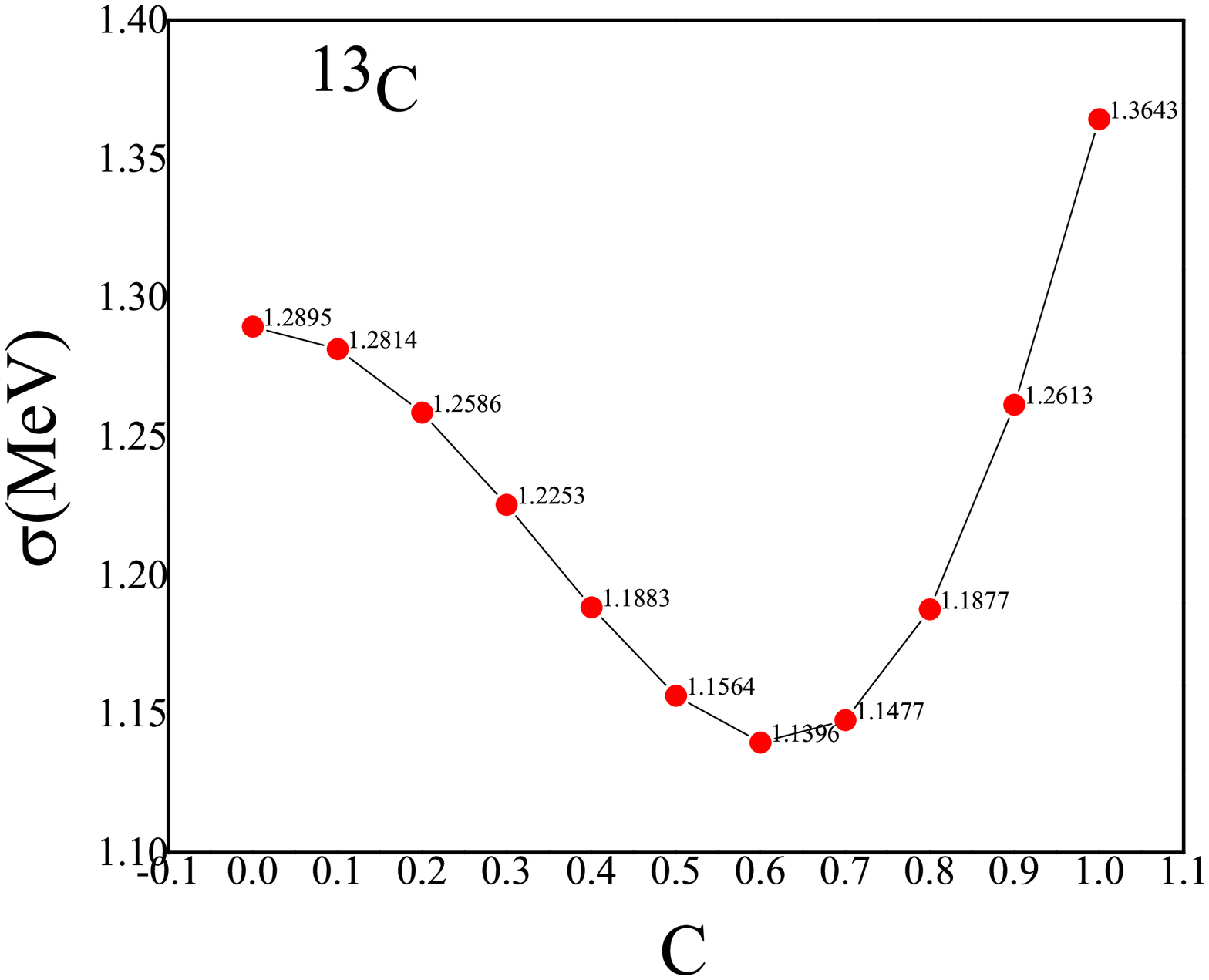}
\includegraphics[height=6cm]{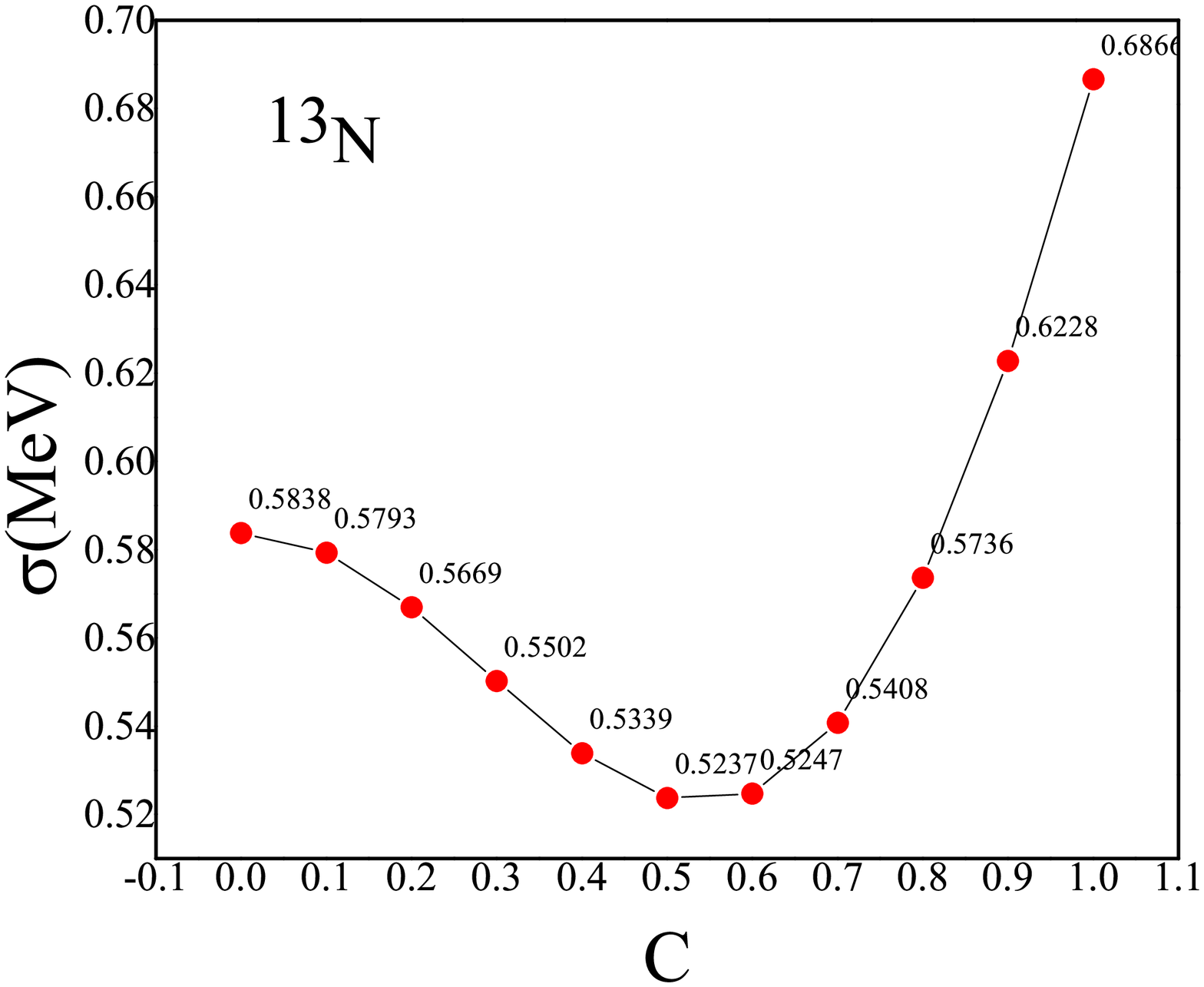}
\includegraphics[height=6cm]{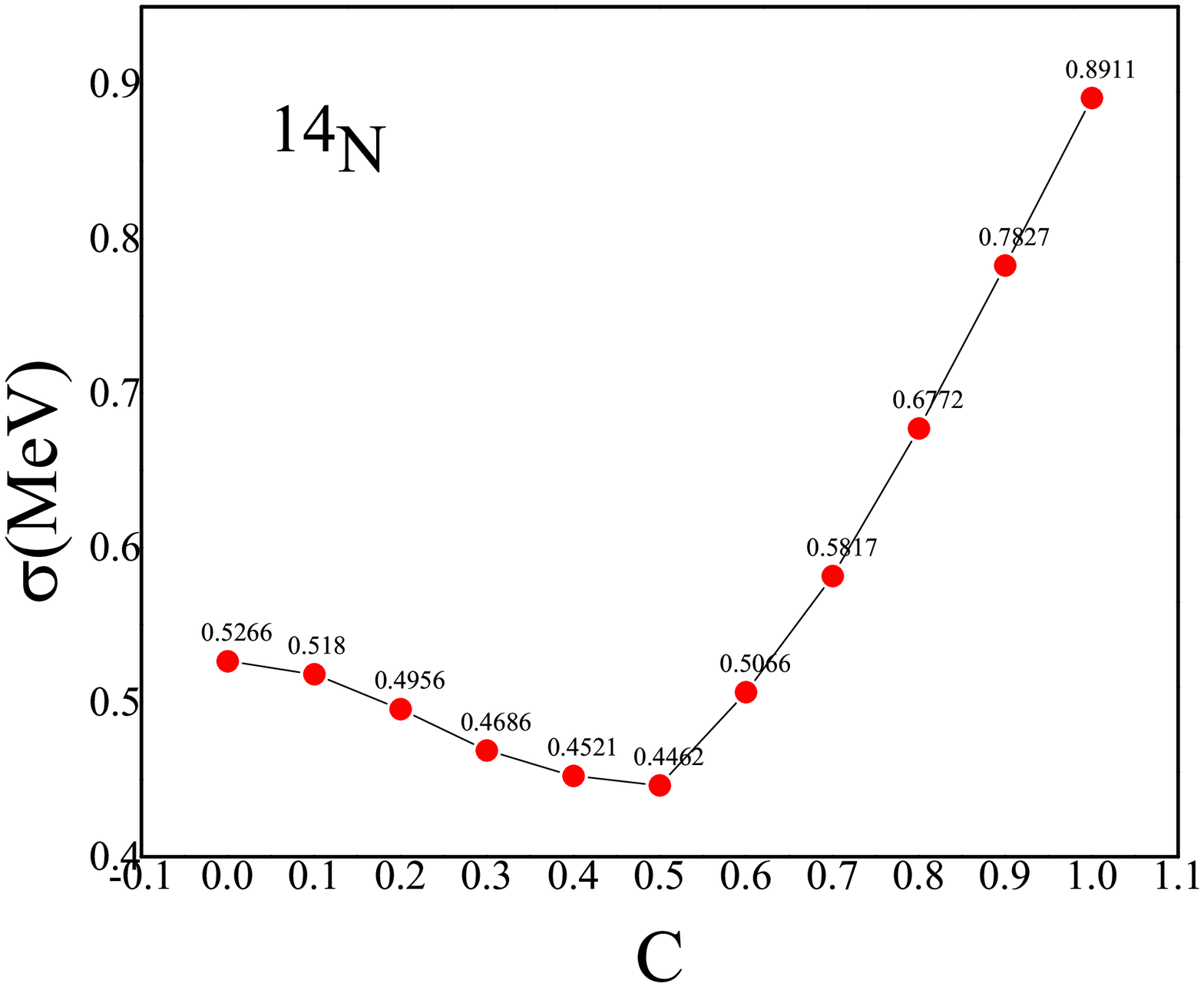}
\caption{ The root mean square deviation as a function of the control parameter C for three-cluster nuclei.}\label{fig:5}
\end{center}
\end{figure}

\begin{figure}
\begin{center}
\includegraphics[height=6cm]{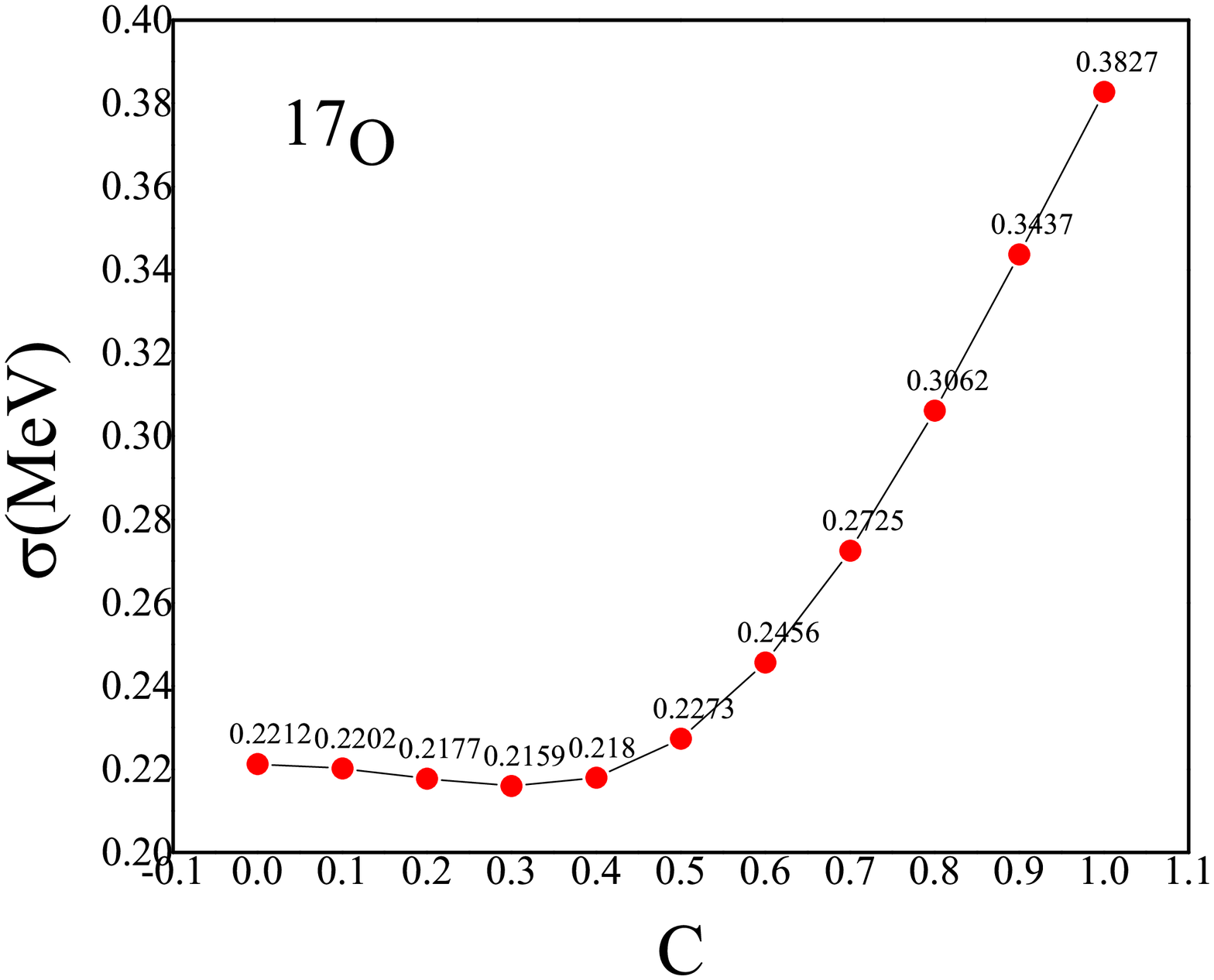}
\includegraphics[height=6cm]{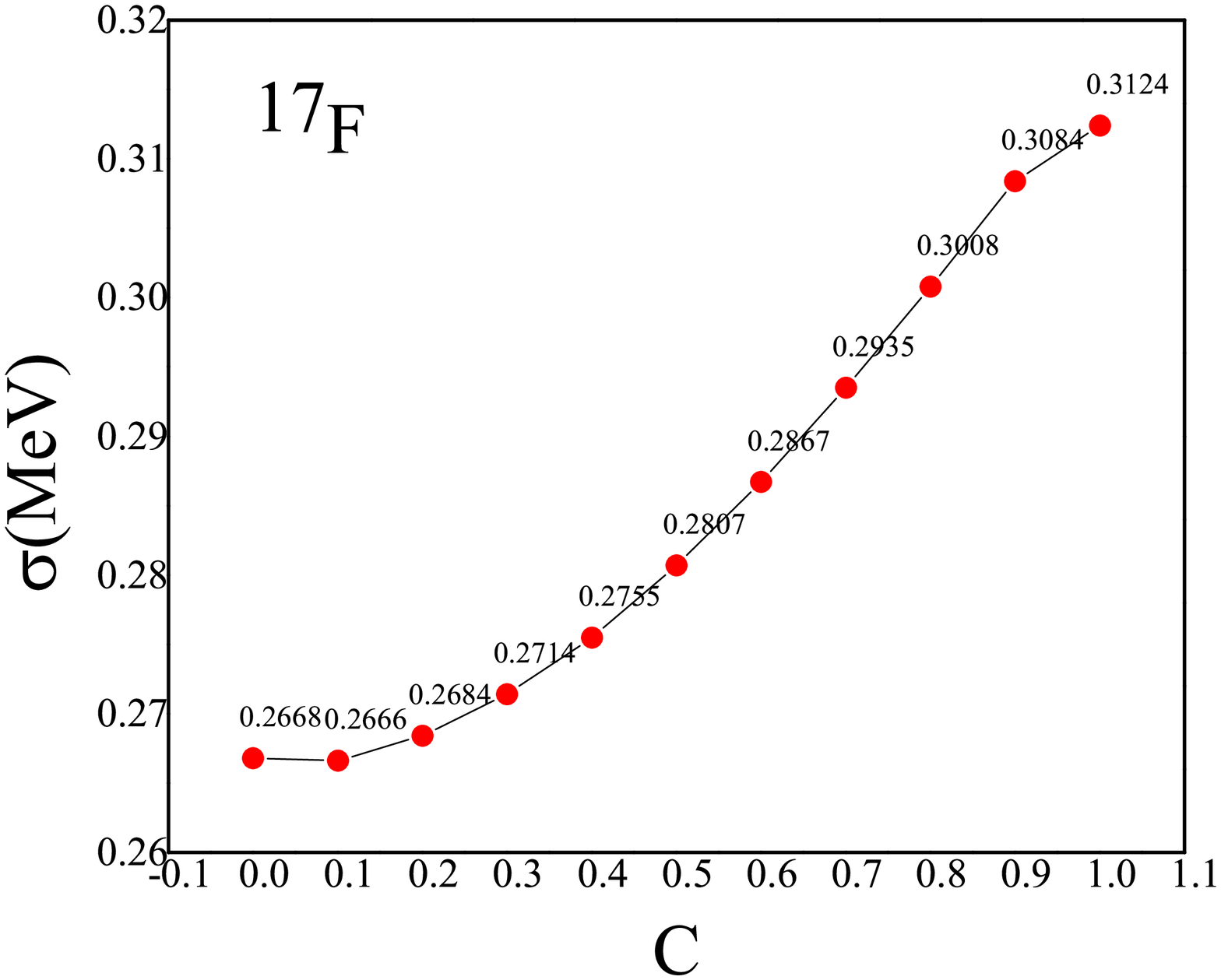}
\caption{ The root mean square deviation as a function of the control parameter C for four-cluster nuclei.}\label{fig:6}
\end{center}
\end{figure}

\begin{flushleft}
\end{flushleft}

\begin{figure}
\begin{center}
\includegraphics[height=6cm]{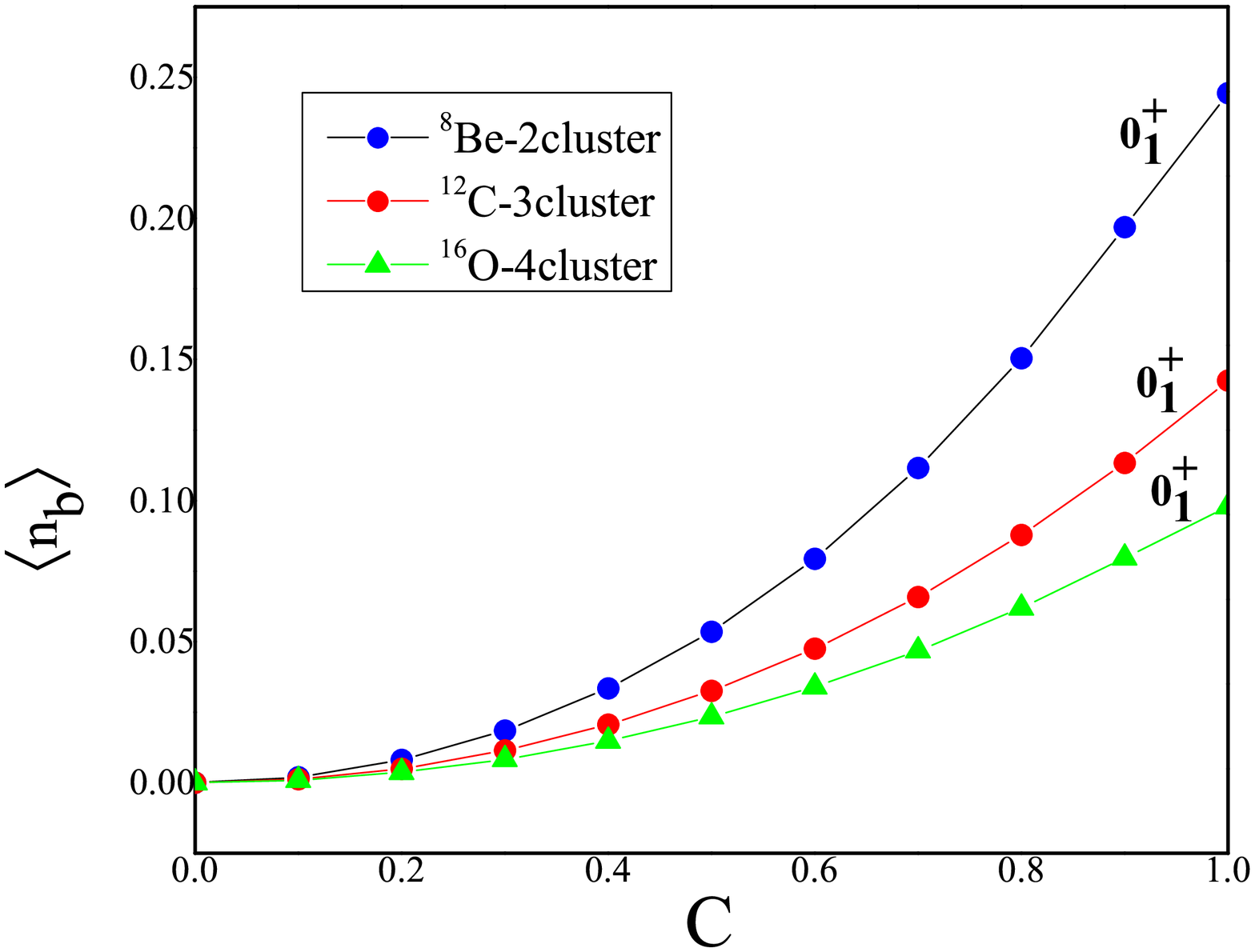}
\includegraphics[height=6cm]{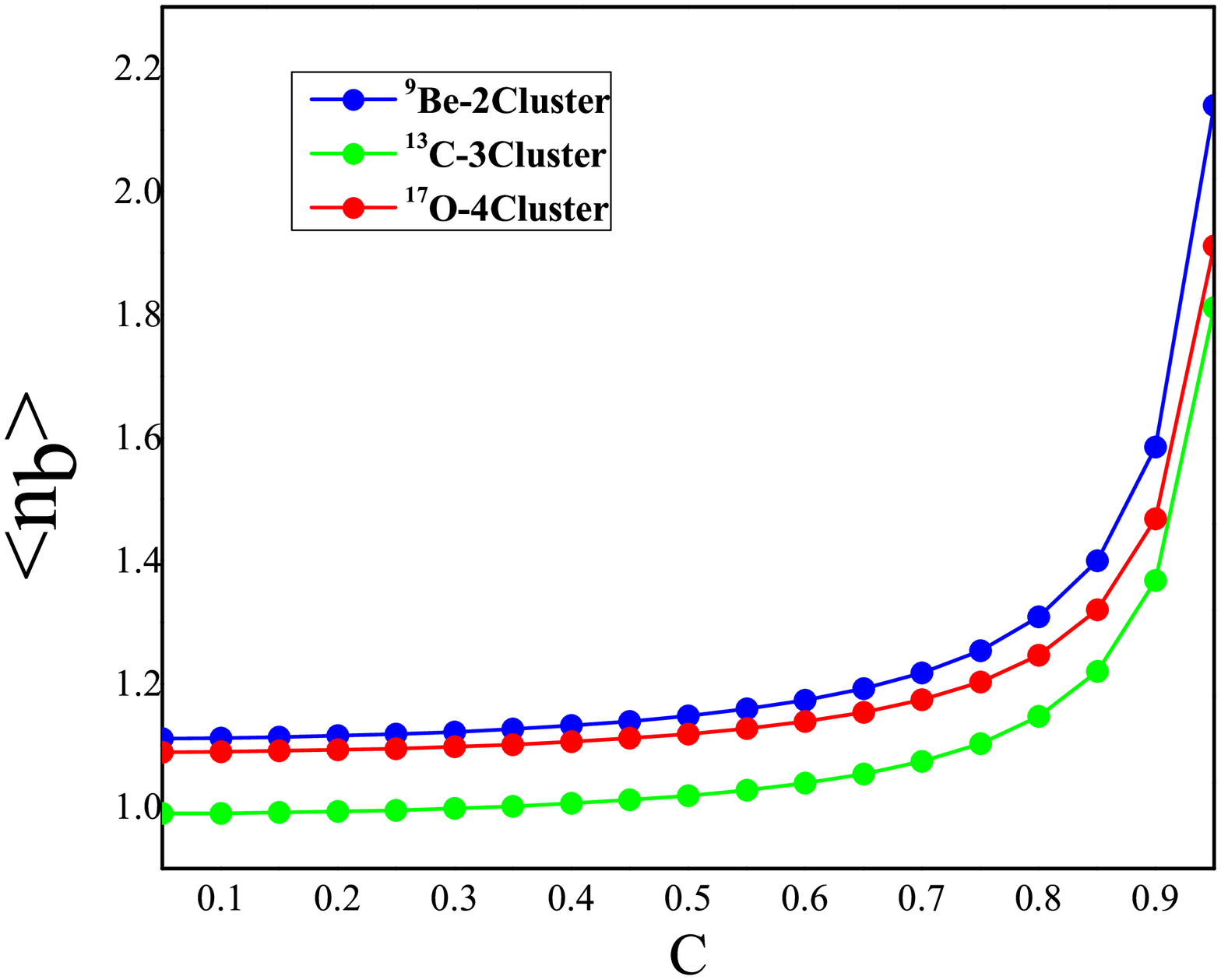}
\caption{The expectation values of the vector-boson number operator for the lowest states as a function of control parameter C for N=10.
}\label{fig:8}
\end{center}
\end{figure}

\begin{flushleft}
\end{flushleft}

\begin{figure}
\begin{center}

\includegraphics[height=6cm]{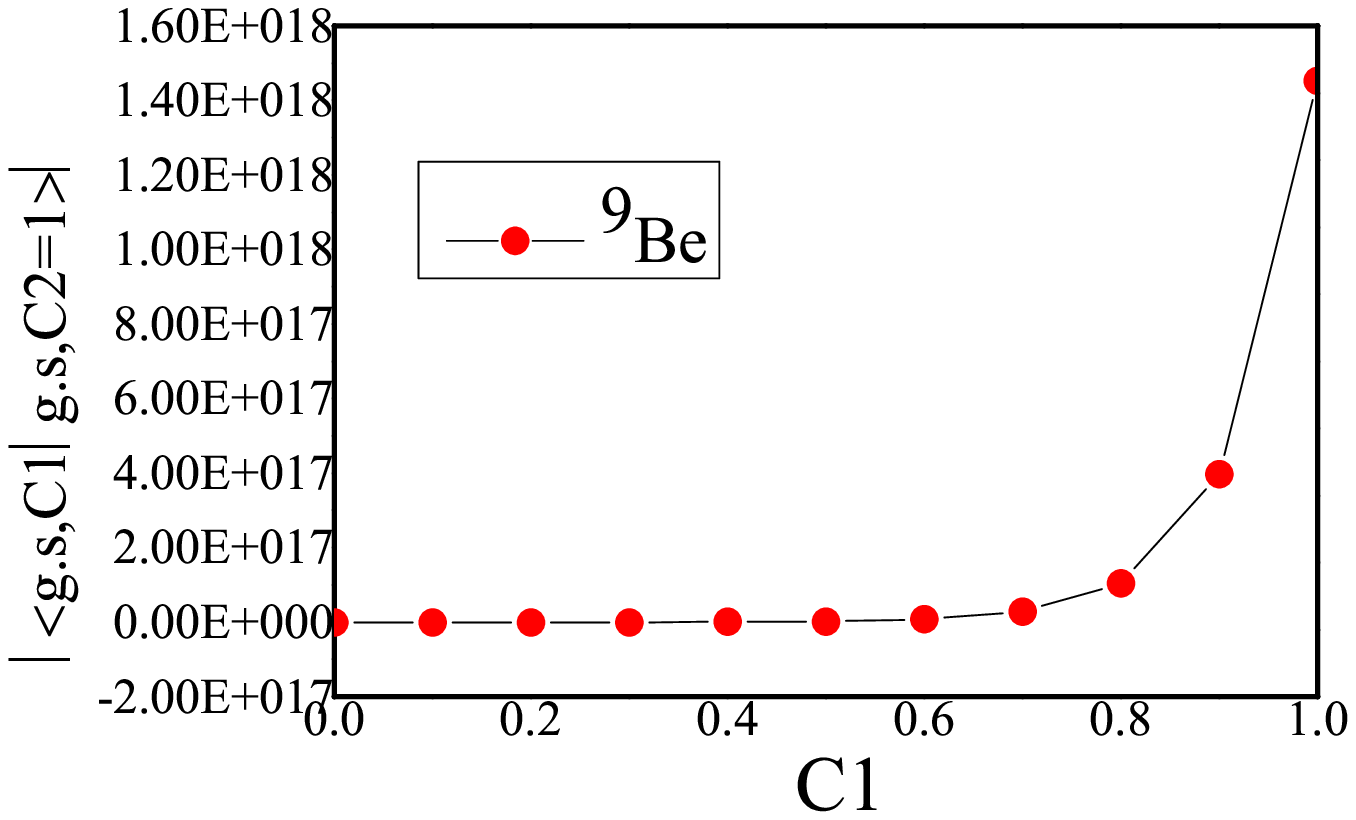}
\includegraphics[height=6cm]{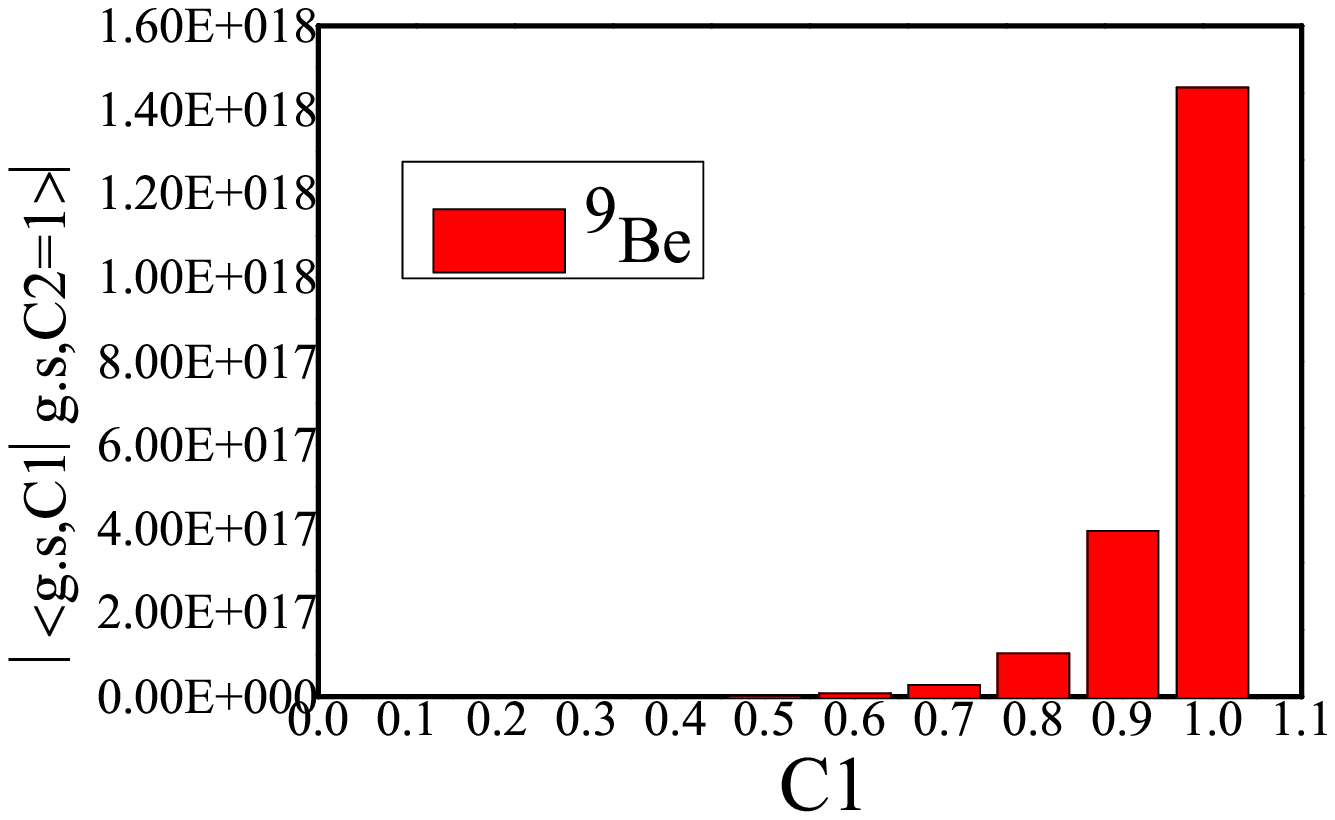}
\includegraphics[height=6cm]{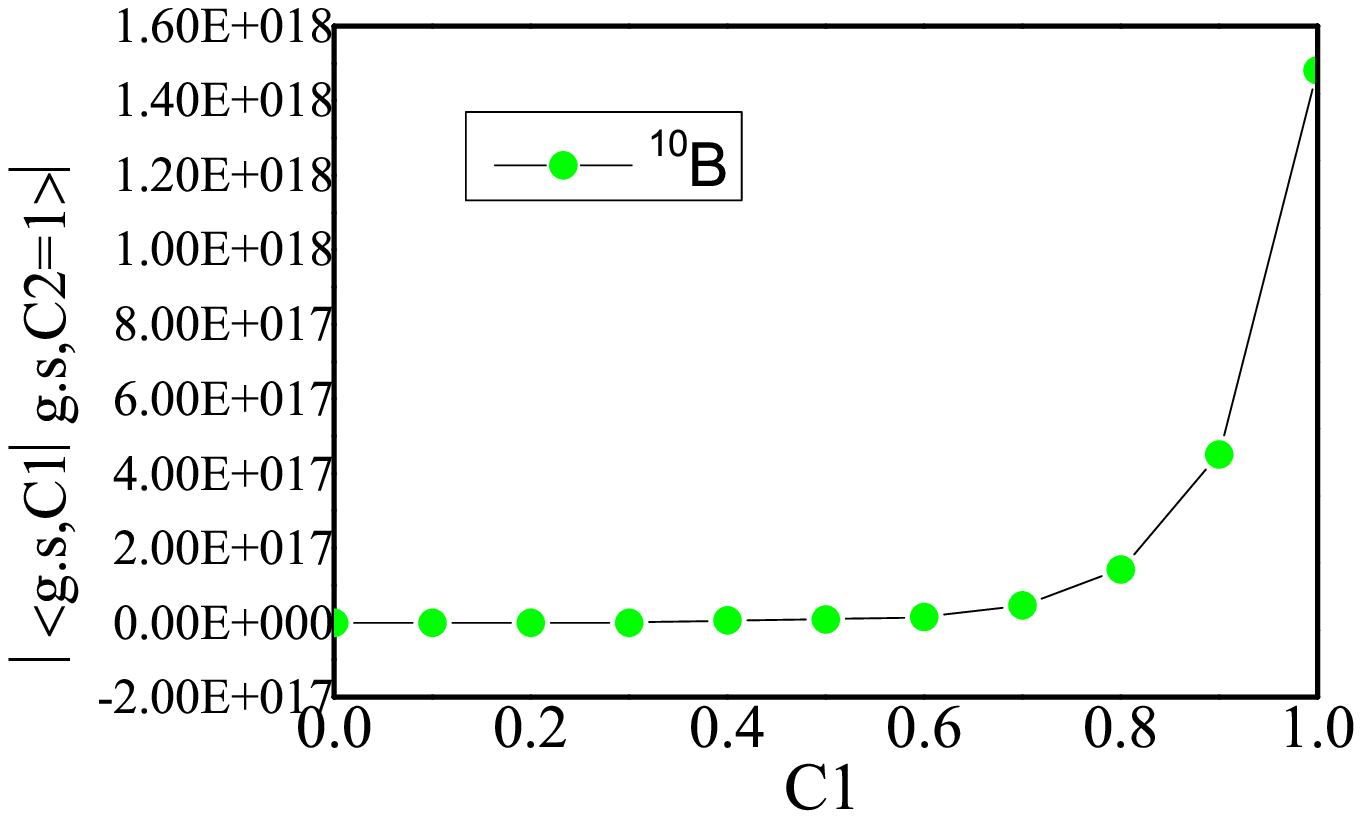}
\includegraphics[height=6cm]{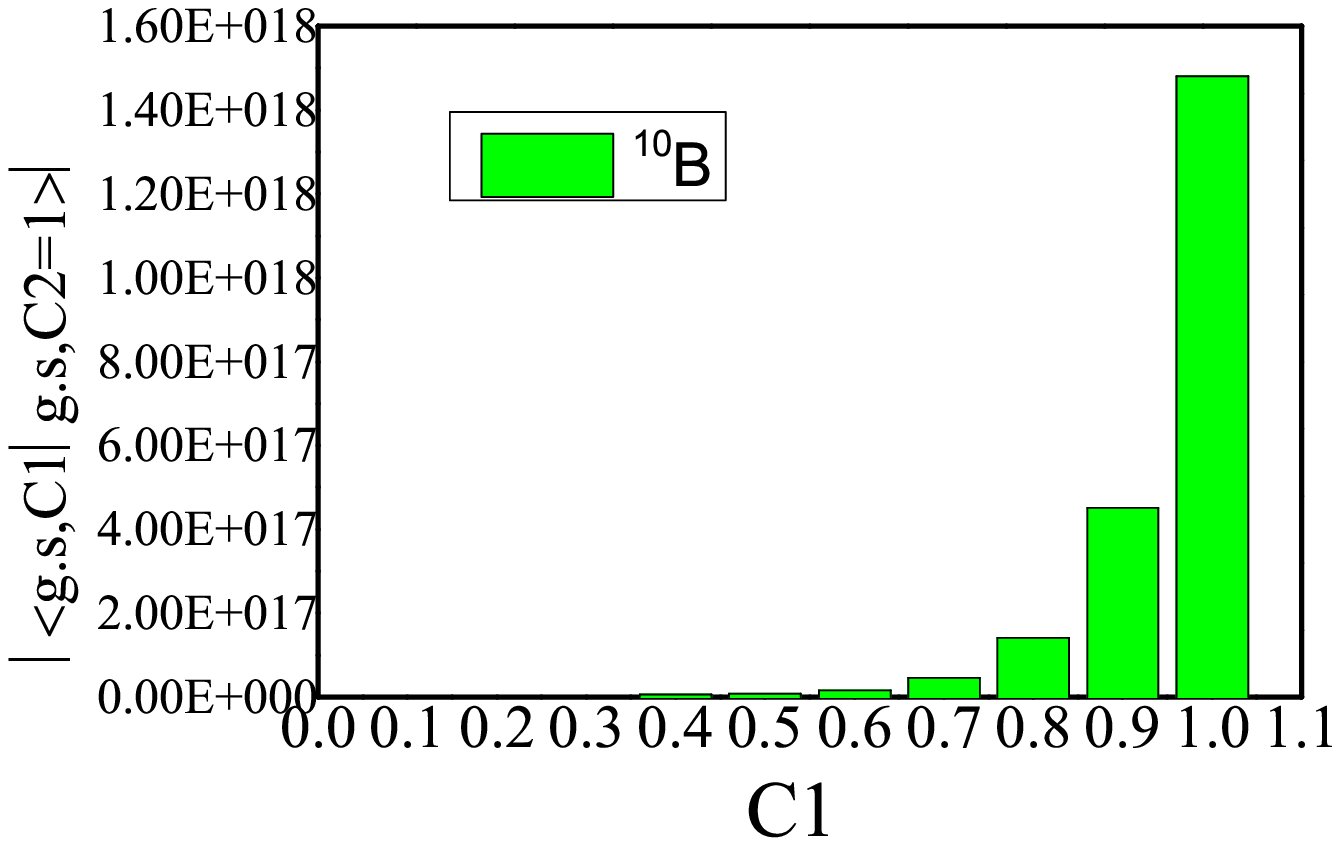}
\caption{The Calculated variation behavior of the overlap of the ground-state wave function as a function of control parameter C for N=10 .
}\label{fig:10}
\end{center}
\end{figure}

\clearpage


\begin{thebibliography}{99}
\bibitem{1} Iachello, F., Iachello, F., Arima, A., and Iachello, F. (1987). The interacting boson model. Cambridge University Press.
\bibitem{2}  Iachello, F., and Levine, R. D. (1995). Algebraic theory of molecules. Oxford University Press.

\bibitem{3}  Bijker, R. (2012). Spectrum generating algebras for few-body systems. In Journal of Physics: Conference Series (Vol. 380, No. 1, p. 012003). IOP Publishing.


\bibitem{4} Bijker, R., and Iachello, F. (2017). The algebraic cluster model: Structure of ${}^{16}O$. Nuclear Physics A, 957, 154-176.
\bibitem{5} Bijker, R., and Iachello, F. (2000). Cluster states in nuclei as representations of a $ U (\nu+ 1) $ group. Physical Review C, 61(6), 067305.
\bibitem{6} Bijker, R., and Iachello, F. (2002). The algebraic cluster model: Three-body clusters. Annals of Physics, 298(2), 334-360.

\bibitem{7} Iachello, F., and Jackson, A. D. (1982). A phenomenological approch to $ \alpha $-clustering in heavy nuclei. Physics Letters B, 108(3), 151-154.
\bibitem{8} Iachello, F. (1983). Algebraic approach to nuclear structure. Nuclear Physics A, 396, 233-243.
\bibitem{9} Daley, H., and Iachello, F. (1983). Alpha-clustering in heavy nuclei. Physics Letters B, 131(4-6), 281-284.
\bibitem{10} Iachello, F., Mukhopadhyay, N. C., and Zhang, L. (1991). Spectrum-generating algebra for stringlike mesons: Mass formula for
$q\bar q$  mesons. Physical Review D, 44(3), 898.
\bibitem{11}  Iachello, F., and Kusnezov, D. (1992). Radiative decays of $(Q\bar Q)$  mesons. Physical Review D, 45(11), 4156.
\bibitem{12} Bijker, R., Iachello, F., and Leviatan, A. (1994). Algebraic models of hadron structure. I. Nonstrange baryons. Annals of Physics, 236(1), 69-116.
\bibitem{13} Bijker, R., Iachello, F., and Leviatan, A. (2000). Algebraic models of hadron structure: II. Strange baryons. Annals of Physics, 284(1), 89-133.
\bibitem{14} Bijker, R., Dieperink, A. E. L., and Leviatan, A. (1995). Spectrum-generating algebra for $ X_3 $ molecules. Physical Review A, 52(4), 2786.
\bibitem{15} Bijker, R., and Leviatan, A. (1998). Algebraic treatment of three-body problems. Few-Body Systems, 25(1-3), 89-10
\bibitem{16}  Bijker, R., and Iachello, F. (2014). Evidence for Tetrahedral Symmetry in ${}^{16}O$. Physical review letters, 112(15), 152501.
\bibitem{17} Majarshin, A. J., Sabri, H., Jafarizadeh, M. A. and  Amiri, N. (2018). Energy spectra of vibron and cluster models in molecular and nuclear systems. The European Physical Journal A, 54(3), 36.
\bibitem{18} von Oertzen, W., Freer, M., and Kanada-En’yo, Y. (2006). Nuclear clusters and nuclear molecules. Physics Reports, 432(2), 43-113.
\bibitem{19}  Hiura, J., and Shimodaya, I. (1963). Alpha-Particle Model for ${}^9Be$. Progress of Theoretical Physics, 30(5), 585-600.
\bibitem{20} Abe, Y., Hiura, J., and Tanaka, H. (1973). A molecular-orbital model of the atomic nuclei. Progress of Theoretical Physics, 49(3), 800-824.
\bibitem{21} Okabe, S., Abe, Y., and Tanaka, H. (1977). The Structure of ${}^9Be$ Nucleus by a Molecular Model. I. Progress of Theoretical Physics, 57(3), 866-881.
\bibitem{22}  Okabe, S., and Abe, Y. (1979). The Structure of ${}^9Be$ by a Molecular Model. II. Progress of Theoretical Physics, 61(4), 1049-1064.
\bibitem{23} Feldmeier, H., and Schnack, J. (2000). Molecular dynamics for fermions. Reviews of Modern Physics, 72(3), 655.
\bibitem{24}  Roth, R., Neff, T., Hergert, H., and Feldmeier, H. (2004). Nuclear structure based on correlated realistic nucleon–nucleon potentials. Nuclear Physics A, 745(1-2), 3-33.
\bibitem{25}  Neff, T., and Feldmeier, H. (2003). Cluster structures within fermionic molecular dynamics. arXiv preprint nucl-th/0312130.
\bibitem{26}  Neff, T., Feldmeier, H., and Roth, R. (2005). Structure of light nuclei in Fermionic Molecular Dynamics. Nuclear Physics A, 752, 321-324.
\bibitem{27}  Kanada-En'yo, Y., and Horiuchi, H. (2001). Structure of light unstable nuclei studied with antisymmetrized molecular dynamics. Progress of Theoretical Physics Supplement, 142, 205-263.
\bibitem{28}  Kanada-En'yo, Y., Kimura, M., and Horiuchi, H. (2003). Antisymmetrized Molecular Dynamics: a new insight into the structure of nuclei. Comptes Rendus Physique, 4(4-5), 497-520.
\bibitem{29}  Kanada-En’yo, Y., and Horiuchi, H. (2003). Cluster structures of the ground and excited states of ${}^{12}Be$studied with antisymmetrized molecular dynamics. Physical Review C, 68(1), 014319.
\bibitem{30}  Brink, D. M., Friedrich, H., Weiguny, A., and Wong, C. W. (1970). Investigation of the alpha-particle model for light nuclei. Physics Letters B, 33(2), 143-146.
\bibitem{31}  Della Rocca, V., Bijker, R., and Iachello, F. (2017). Single-particle levels in cluster potentials. Nuclear Physics A, 966, 158-184.
\bibitem{32} Bijker, R., and Iachello, F. (2019). Evidence for Triangular  $ D_{3h} $  Symmetry in $ {}^{13}C $  
. Physical review letters, 122(16), 162501.
\bibitem{33} Leviatan, A., and Kirson, M. W. (1988). Intrinsic and collective structure of an algebraic model of molecular rotation-vibration spectra. Annals of Physics, 188(1), 142-185.
\bibitem{34} Roosmalen, O. S. V. (1982). Algebraic descriptions of nuclear and molecular rotation-vibration spectra (Doctoral dissertation, Rijksuniversiteit Groningen).
\bibitem{35}  Iachello, F., and Levine, R. D. (1982). Algebraic approach to molecular rotation‐vibration spectra. I. Diatomic molecules. The Journal of Chemical Physics, 77(6), 3046-3055.
\bibitem{36} Bijker, R. (2010, December). Algebraic cluster model with tetrahedral symmetry. In AIP Conference Proceedings (Vol. 1323, No. 1, pp. 28-39). AIP.




\bibitem{37} Kramer, P., and Moshinsky, M. (1966). Group theory of harmonic oscillators (III). States with permutational symmetry. Nuclear Physics, 82(2), 241-274.
\bibitem{38}  Pan, F., and Draayer, J. P. (1998). New algebraic solutions for $ SO (6)\leftrightarrow U (5) $ transitional nuclei in the interacting boson model. Nuclear Physics A, 636(2), 156-168.
\bibitem{39}  Pan, F., Zhang, X., and Draayer, J. P. (2002). Algebraic solutions of an sl-boson system in the $ U (2l+ 1)\leftrightarrow O (2l+ 2) $ transitional region. Journal of Physics A: Mathematical and General, 35(33), 7173.
\bibitem{40}  Iachello, F., and Arima, A. (1974). Boson symmetries in vibrational nuclei. Physics Letters B, 53(4), 309-312.
\bibitem{41} Tilley, D. R., Kelley, J. H., Godwin, J. L., Millener, D. J., Purcell, J. E., Sheu, C. G., and Weller, H. R. (2004). Energy levels of light nuclei A= 8, 9, 10. Nuclear Physics A, 745(3-4), 155-362.
\bibitem{42} Ajzenberg-Selove, F. (1979). Energy levels of light nuclei
$A = 5 - 10$. Nuclear Physics A, 320(1), 1-224.
\bibitem{43}  Ajzenberg-Selove, F. (1984). Energy levels of light nuclei $A = 5 - 10$. Nuclear Physics A, 413(1), 1-168.
\bibitem{44} Jolie, J., Cejnar, P., Casten, R. F., Heinze, S., Linnemann, A., and Werner, V. (2002). Triple point of nuclear deformations. Physical review letters, 89(18), 182502.
\bibitem{45} Pan, F., Zhang, Y., and Draayer, J. P. (2005). Quantum phase transitions in the $ U (5)\leftrightarrow O (6) $ large-N limit. Journal of Physics G: Nuclear and Particle Physics, 31(9), 1039.
\bibitem{46} Zhang, Y., Hou, Z. F., and Liu, Y. X. (2007). Distinguishing a first order from a second order nuclear shape phase transition in the interacting boson model. Physical Review C, 76(1), 011305.
\bibitem{47} Zhang, Y., Hou, Z. F., Chen, H., Wei, H., and Liu, Y. X. (2008). Quantum phase transition in the $ U(4) $ vibron model and the $ E(3) $ symmetry. Physical Review C, 78(2), 024314. 



\end{thebibliography}
\end{document}